\documentclass{acm_proc_article-sp}
\usepackage[utf8]{inputenc}
\usepackage[numbers,sort]{natbib}
\pdfoutput=1
\usepackage{tikz}
\usetikzlibrary{shapes,arrows}

\usepackage[english]{babel}
\usepackage{url}
\usepackage{balance}
\usepackage{graphicx}
\usepackage{caption}
\DeclareCaptionType{copyrightbox}
\usepackage{subcaption}
\usepackage{multirow}
\usepackage{todonotes}

\usepackage{setspace}
\DeclareMathOperator*{\argmin}{arg\,min}

\usepackage{color}
\usepackage{soul}
\newcommand{\hilight}[1]{#1}
\title{
Can't you hear me knocking: \\Identification of user actions on Android apps\\via traffic analysis
}

\numberofauthors{4} %  in this sample file, there are a *total*
\author{
\alignauthor Mauro~Conti\\ 
\affaddr{University of Padua}\\
\affaddr{Padua, Italy}\\
\email{conti@math.unipd.it}
\alignauthor Luigi~V.~Mancini\\
       \affaddr{Sapienza University of Rome}\\
       \affaddr{Rome, Italy}\\
       \email{lv.mancini@di.uniroma1.it}
\alignauthor Riccardo Spolaor\\
       \affaddr{University of Padua}\\
       \affaddr{Padua, Italy}\\
       \email{spolaor.riccado@gmail.com}
\and  % use '\and' if you need 'another row' of author names
\alignauthor Nino~V.~Verde\\
       \affaddr{Sapienza University of Rome}\\
       \affaddr{Rome, Italy}\\
       \email{verde@di.uniroma1.it}
}
\begin{document}

\maketitle

\begin{abstract}
While smartphone usage become more and more pervasive, people start also asking to which extent
such devices can be maliciously exploited as ``tracking devices''. The concern is not only related
to an adversary taking physical or remote control of the device (e.g., via a malicious app),
but also to what a passive adversary (without the above capabilities) can observe from the device communications.
% can be observed in a passive way, simply by eavesdropping on the phone communications.
Work in this latter direction aimed, for example, at inferring the apps a user has installed
on his device, or identifying the presence of a specific user within a network.

%NB we consider only user action done by a user (sending email/tweet/post), NOT from other users interactions (receiving email)
In this paper, we move a step forward: we investigate to which extent it is feasible to identify the specific
actions that a user is doing on his mobile device, by simply eavesdropping the device's network traffic.
In particular, we aim at identifying actions like \hilight{browsing someone's profile on a social network, posting a message on a
friend's wall, or sending an email.}
We design a system that achieves this goal starting from encrypted TCP/IP packets: it works through identification of network 
flows and application of machine learning techniques. We did a complete implementation of this system and
run a thorough set of experiments, which show that it can achieve accuracy and precision higher than 95\%,
for most of the considered actions.
\end{abstract}

%%%%%%%%%%%%%%%%%%%%%%%%%%%%%%%%%%%%%%%%%%%%%%%%%%%%%%%%%%%%%%%%%%
\section{Introduction}
\label{Introduction}

Smartphones become widely used and pervasive devices. People continuously carry those devices with them and use them more and more for daily communication activities, including not only voice calls and SMS but also emails and social network interaction.
In the last years, several concerns have been raised about the capabilities of those portable devices to invade the privacy of the users and actually becoming ``tracking devices''. In particular, one aspect is concerned with the possibility of continuously localize an individual \cite{iphonetrackposition:14,angelaTrack:14}.
Another relevant aspect is related to the fact that malicious apps can go even a step further in tracing and spying on someone life. For example, a malicious app that has access to the microphone and networking capabilities, could in principle continuously eavesdrop the audio and send it over the Internet to an adversary \cite{schlegel2011soundcomber}. 

Even when the adversary has no actual control of the phone (either physical control or remote via malicious apps) other attacks in the same directions are possible to violate the privacy of the communications. 
If the network traffic is not encrypted, the task of the eavesdropper is simple, since he can analyze the payload and read the content of each packet. However, many mobile apps use the \hilight{Secure Sockets Layer (SSL), and its successor Transport Layer  Security (TLS),} as a building block for encrypted communications.
In a typical SSL/TLS usage scenario, a server is configured with a certificate 
containing a public key as well as a matching private key. As part of the 
handshake between an SSL/TLS client and server, the server proves it has the 
private key by signing its certificate with public-key cryptography. 
Unfortunately there is often a gap between theory and practice, e.g., leveraging the SSL vulnerabilities of smartphone apps \cite{Fahl:2012:WEM:2382196.2382205,Georgiev:2012:MDC:2382196.2382204} one might run an SSL man-in-the-middle attack to compromise the confidentiality of communications.

We believe that while people become more familiar with mobile technologies and their related privacy threats (also thanks to the attention raised by the media, e.g., see the recent attention on NSA for supposedly eavesdropping foreign governments leader such as Angela Merkel \cite{angelaTrack:14}), users start adopting some good practices that better adapt to their privacy feeling and understanding. For examples, solutions to identify and isolate malware running on smartphones \cite{InfFlow-TIFS,MOSES-TPDS,survey13suarez} as well as to protect against attacks coming from the network \cite{DBLP:ContiDG13mithys,LeoneConti.6732964} might significantly reduce current threats to user privacy.

Unfortunately, we believe that even adopting such good practices would not close the door to malicious adversaries willing to trace people. In fact, the wireless and pervasive nature of mobile devices would still leave many practical options for adversarial tracing. 
In particular, even when such solutions are in place, the adversary can still infer a significant amount of information from the properly encrypted traffic.
For example, work leveraging analysis of encrypted traffic already highlighted the possibility of understanding the apps a user has installed on his device \cite{Stober:2013:YSY:2462096.2462099}, or identify the presence of a specific user within a network \cite{NinoVerdeNATleftBehind}.

%\newpage
This work focuses on understanding whether the user profiling made through analyzing encrypted traffic can be pushed up to understand exactly what actions the user is doing on his phone: as concrete examples, we aim at identifying actions such as the user sending an email, receiving an email, browsing someone profile in a social network, \hilight{rather than publishing a post or a tweet.} %``tagging" someone in a picture.
The underlying issue we leverage in our work is that SSL and TLS protect the content of a packet, while they do not 
prevent the detection of networks packets patterns that instead may reveal some information about the user behavior.

%\paragraph{Motivations}
\hilight{
%convincere il reviewer 
An adversary may use our approach in several practical ways to threaten the privacy of the user.
In the following, we report some possible attacks:}
%We believe that an adversary can use our approach in several practical ways to threaten the privacy of the user.
%We report in the following some possible attacks:}
%Indeed, a common user is not aware that information disclosure about his actions can be such a relevant threat to his privacy.} %For example, the adversary can learn 
%the victim habits and 
%In this paragraph, we report some possible scenarios where an adversary uses this knowledge.
\begin{itemize}
\item \hilight{
A censorship government may try to identify a dissident who spreads anti-government propaganda 
using an anonymous social network account. Comparing the time of the public posts with the time of the actions
(inferred with our method), the government can guess the identity of that anonymous dissident.}
%Let us consider a censorship government that is trying to identify a dissident who spreads anti-government propaganda, using an anonymous social network account. Comparing the time of the posts with the time of the actions (inferred with our method), the government can guess the identity of that anonymous dissident.}
\item \hilight{By tracing the actions performed by two users, and taking into account the 
communication latency, an adversary may guess (even if with some probability of error) whether there is a communication 
between them. Multiple observations could reduce the probability of errors.}
\item \hilight{An adversary can build a behavioral profile of a target victim based on the habits of the latter one
(e.g., wake up time, work time).
%Being the behavior unique for a user, it is possible to trace his movements, even if he changes his mobile device.
For example, this could be used to improve user fingerprinting methods, to infer the presence of a 
particular user in a network}~\cite{NinoVerdeNATleftBehind}, \hilight{even when he accesses the network 
with different type of devices.}

%\item \hilight{A company that provides Internet access (an ISP such as AT\&T, or just as free hot-spot provider as Starbucks or McDonald) can use the information inferred with our proposed method to gain further commercial or intelligence advantage.}
%\item \hilight{Let us assume that an adversary aims to read a private message sent by a user. Having the knowledge of the precise time when that private message was sent, the adversary can  focus his efforts trying to decrypt the traffic in that particular time frame. An adversary without that knowledge has to focus his efforts on the whole traffic capture.}
\end{itemize}

\paragraph{Contributions}
In this paper, we propose a framework to infer which particular actions the 
user executed on some app installed on his mobile-phone, by only looking 
at the network traffic that the phone generates. In particular, we assume the traffic is encrypted and the adversary eavesdrops (without modifying them) the messages exchanged between the user's device and the web services that he uses. 

Our framework analyzes the network communications and leverages information available 
in TCP/IP packets (like IP addresses and ports), together with other 
information like the size, direction (incoming/outgoing), and timing. By 
using an approach based on machine learning, each app that is of interest is analyzed independently. 
To set up our system, for each app we first pre-process a dataset of network packets labeled with 
the user actions that originated them, we cluster them in flow typologies that represent recurrent network flows, and 
finally  we analyze them in order to create a training set that will be used to feed a 
classifier. The trained classifier will be then able to classify new traffic 
traces that have never been seen before.
We fully implemented our system, %which we called Angela (to recall the Angela Merkle case \cite{angelaTrack:14}), 
and we run a thorough set of experiments to evaluate our solution considering three very popular apps: Facebook, Gmail, and Twitter.
%In the following, we will detail the process applied to each of the apps analyzed.
The results shows that it can achieve accuracy and precision higher than 95\%, for most of the considered actions done by the user with those apps.

 \paragraph{Organization}
The remainder of the paper is organized as follows. 
In Section \ref{Related Work}, we revise the state of the art around our research topic. 
In Section \ref{ML and DM tools}, we introduce some background knowledge, used in our work, on machine learning and data mining tools.
In Section \ref{OurFramework}, we present our framework, describing it in all its components.
We present the evaluation of our solution for identifying user actions in Section \ref{ExperimentalResults}, 
while in Section \ref{Countermeasures} we discuss about possible countermeasures against the attack.
Finally, in Section \ref{Conclusions} we draw some conclusions.

%%%%%%%%%%%%%%%%%%%%%%%%%%%%%%%%%%%%%%%%%%%%%%%%%%%%%%%%%%%%%%%%%%%
\section{Related Work}
\label{Related Work}

Our main claim in this paper is that network traffic analysis and machine learning can be used to infer
private information about the user, i.e., the actions that he executes with his mobile phone,
even thought the traffic is encrypted.
To position our contribution with respect to the state of the art, in this section we survey the works that belong to two main research areas that focus on  
similar issues: \textit{privacy attacks via traffic analysis} (not necessarily focusing on mobile devices) and \textit{traffic analysis of mobile devices} (not necessarily focusing on privacy).

\paragraph{Privacy attacks via traffic analysis}

In the literature, several works proposed to track user activities on the web 
by analyzing unencrypted HTTP requests and responses~\cite{Atterer:2006:KUM:1135777.1135811,Schneider:2009:UOS:1644893.1644899,Benevenuto:2012:CUN:2169463.2169583}.
With this analysis it was possible to understand user actions inferring interests and habits. 
%In  it is study how users interact with Online Social Network (OSN) analysing anonymized HTTP header.\cite{Schneider:2009:UOS:1644893.1644899}
%With their method, It is possible Following user activity on OSN web pages, it can be infere user actions .\\
%Characterizing user navigation and interactions in online social networks \cite{Benevenuto:2012:CUN:2169463.2169583}
However, in recent years, websites and social networks started to use SSL/TLS encryption protocol, 
both for web and mobile services. %~\cite{FacebookSSL:13,APITwitterSSL:14}. 
This means that communications between endpoints are encrypted and this type of analysis cannot be performed anymore. 

Different works surveyed possible attacks that can be performed using traffic analysis assuming a 
very strong adversary (e.g., a national security agency) which is able to observe all communication 
links \cite{Raymond:2001:TAP:371931.371972,Berthold:2000:PLU:332186.332211}.
In~\cite{Liberatore:2006:ISE:1180405.1180437}, Liberatore et al.\ evaluated the effectiveness of two traffic analysis 
techniques based on naive Bayes and on Jaccard’s coefficient for identifying encrypted HTTP streams. 
Such an attack was outperformed by~\cite{Herrmann:2009:WFA:1655008.1655013}, where the authors presented 
a method that applies common text mining techniques to the normalized frequency distribution of observable 
IP packet sizes, obtaining a classifier that correctly identifies up to 97\% of requests.
The technique was further refined in~\cite{Panchenko:2011:WFO:2046556.2046570}, where the authors presented a support 
vector machine classifier that was able to correctly identify web pages, even when the victim used both encryption and 
anonymization networks such as Tor~\cite{Dingledine:2004:TSO:1251375}.
\hilight{Finally, Cai et al.}~\cite{cai2012touching} 
\hilight{ presented a web pages fingerprinting attack and proved its effectiveness despite traffic analysis countermeasures, such as Tor randomized pipelining} ~\cite{tor:websites:fingerprint} or HTTPOS~\cite{Luo11httpos:sealing}.

Unfortunately, none of the aforementioned works was designed for (or could easily be extended) to mobile devices. In fact, all of them focus on web pages identification in desktop environment (in particular, in desktop browsers), where the generated HTTP traffic strictly depends on how web pages are designed. Conversely, mobile users mostly access the contents through the apps installed on their devices~\cite{Go:2013:TAA:2444776.2444779}. %,appmobilebrowser}.
These apps communicate with a service provider (e.g., Facebook) through a set of APIs.
% \hilight{Moreover, Cai et al. in} \cite{cai2012touching} 
% \hilight{present a web pages fingerprint attack and prove its effectiveness despite traffic analysis countermeasures, like randomized pipelining in Tor} \cite{tor:websites:fingerprint} and HTTPOS \cite{Luo11httpos:sealing}.
% \hilight{The works reported before are focused on web pages identification in desktop environment (browser), so the traffic they produce is related to how the web page is designed.
% In the other hand, Android native apps are coded with a specific programming language (Java) and they use libraries provided by Android API.
% So instead of web pages identified by an URL, users browse an app through ''intents`` and graphical objects are organized in ''views``. 
\hilight{An example of such differences between desktop web browsers and mobile apps is the validation of SSL certificates}~\cite{Georgiev:2012:MDC:2382196.2382204,DBLP:ContiDG13mithys}.

Traffic analysis has been applied not only to HTTP but also to other protocols. 
For example, Song et al.~\cite{Song:2001:TAK:1251327.1251352} prove that SSH is not secure. In particular, they show that 
 even very simple statistical techniques suffice to reveal sensitive 
information such as login passwords. More importantly, the authors show that by using 
more advanced statistical techniques on timing information collected from the 
network, the eavesdropper can also learn significant information about what users 
type in SSH sessions. 
%By developing a Hidden Markov Model, they were able to 
%predict key sequences from the inter-keystroke timings. 
SSH is not the only 
protocol that has been target of such attacks. 
Another example is Voice Over IP (VoIP). In particular, in~\cite{Wright:2008:SMY:1397759.1398055}, the authors 
show how the length of encrypted VoIP packets can be used to identify spoken 
phrases of a variable bit rate encoded call. Their work indicates that a profile 
Hidden Markov Model trained using speaker- and phrase-independent data can 
detect the presence of some phrases within encrypted VoIP calls with recall and 
precision exceeding 90\%.

In ~\cite{shuosp2010}, \hilight{the authors show that despite encryption, 
also web applications suffer from side-channel leakages. The system model considered is different from our. 
In particular, their focus is on web applications. On the contrary, we focus on mobile applications. 
More importantly, the authors leverage three fundamental features of web applications: stateful communication; low entropy input; 
significant traffic distinction. 
We believe that in most mobile applications two of these features (stateful communication, low entropy input) 
are not very useful to characterize user actions. In contrast to this work, we adopt a solution that only needs 
information about packet sizes and their order.
%While we do not have comparative experimental results about the above claim, 
%we still believe that reaching the same target (i.e., recognizing user actions)
%considering less information (packet sizes and order only) is a significant improvement over the state of the art. 
}
%
% %%%%% traffic classification for privacy
% In addiction of application recognition, machine learning techniques applied on network traffic analysis, could be used also to monitor user activities.
% This introduce a privacy issue, because an eavedropper could infere personal sensitive information about users. 
% %%%%%%%%%%%%%%%%%%%%%%%%%%%%%%%%%%%%%%%%%%%%%%%%%%%%%%%%%%%%%%%%%%%
% Machine learning methods like HMM (Hidden Markov Models) are normaly used in speech recognition.
% But it is also possible to use HMM to recognize phrases spoken on a VOIP conversation \cite{Wright:2008:SMY:1397759.1398055}, only by timing and features of encrypted traffic.
% %skype comunication traced \cite{Wang:2005:TAP:1102120.1102133}  non c'entra tanto, no traffic analysis, watermark injection \\
% %Timing analysis of keystrokes and timing attacks on SSH \cite{Song:2001:TAK:1251327.1251352}
% Another example of analysis on encrypted traffic, but related to Secure Shell (SSH) is \cite{Song:2001:TAK:1251327.1251352}.
% In this case, authors exploit inter-keystrokes time to recognize the sequence of characters typed by user.   
%%%%%%%%%%%%%%%%%%%%%%%%%%%%%%%%%%%%%%%%%%%%%%%%%%%%%%%%%%%%%%%%%%%
%infering users' activity....
% %svm 
% In \cite{Zhang:2011:IUO:1998412.1998425} Zhang et al. profile applications behavior using traffic analysis,
% they were able to infere which kind of activity a user is doing by eavedropping MAC-layer transmissions for a few minutes.
%%%%%%%%%%%%%%%%%%%%%%%%%%%%%%%%%%%%%%%%%%%%%%%%%%%%%%%%%%%%%%%%%%%
%Understanding Online Social Network Usage from a Network Perspective

\paragraph{Traffic analysis of mobile devices}
%%%%%%%Android apps behavior by traffic analysis
Focusing on mobile devices, traffic analysis has been successfully used to detect 
information leaks \cite{Enck:2010:TIT:1924943.1924971}, 
to profile users by their set of installed apps~\cite{Stober:2013:YSY:2462096.2462099}, 
\hilight{and to produce the fingerprint of an app from its unencrypted HTTP traffic} \cite{DaiTWNS:2013}.
Traffic analysis has also been used to understand network traffic characteristics, 
with particular attention on energy saving \cite{Falaki:2010:FLT:1879141.1879176}. %Baghel:6176903
%Signature generation for sensitive information leakage in android applications \cite{Kuzuno:6547438}
%%%%%%%%%%%%%%%%%%%%%%%%%%%%%%%%%%%%%%%%%%%%%%%%%%%%%%%%%%%%%%%%%%%
Stober et al. \cite{Stober:2013:YSY:2462096.2462099} shown that it is possible to identify the set of apps installed on an Android device, 
by eavesdropping the 3G/UMTS traffic that those apps generate.
%%%%%%%%%%%%%%%%%%%%%%%%%%%%%%%%%%%%%%%%%%%%%%%%%%%%%%%%%%%%%%%%%%%%%%%%%%%%%%%
%NetworkProfiler: Towards Automatic Fingerprinting of Android Apps
Similarly, Tongaonkar et al. \cite{DaiTWNS:2013} \hilight{introduced an automatic app profiler that creates the network fingerprint
of an Android app relying on packet payload inspection, in order to re-identify its HTTP traffic.} 
% Their framework first collects URLs of HTTP requests from unencrypted payloads
% of network packets generated by app, then builds a fingerprint.
%Understanding mobile app usage patterns using in-app advertisements 
%To increase the accuracy of the profiler, this work has been extended by considering in-app advertisements traffic \cite{Tongaonkar:2013:UMA:2482362.2482371}, while the authors do not deal with encrypted network traffic.
%
%CCS13 TODO
In~\cite{zhouccs2013}, \hilight{Zhou et al. present a work that take in consideration the traffic produced a user action performed with Twitter app.
Unfortunately, the authors focused on a single user action (i.e., send a tweet) without distinguish that action from the other ones a user could perform.}

None of the works mentioned in this section aim at \hilight{inferring and distinguish user actions performed by the user with his mobile apps},
which is the goal of our paper.
% 
% Many of these works claim that In these works, 
% 
% it is also claimed that over 70\% of Android apps traffic is unencrypted \cite{Wei:2012:PMP:2348543.2348563} %ProfileDroid: multi-layer profiling of android applications
% or that encryption protocols aren't used in a proper way \cite{Fahl:2012:WEM:2382196.2382205,Georgiev:2012:MDC:2382196.2382204}.
%Unfortunately these methods couldn't be applied on Android apps we consider in our work.
% It has to be noticed that in captures done for our work, we observed that network traffic produced 
% by Facebook, Gmail and Twitter involves encrypted connections,
% where exchanged packets' protocol is TSLv1 for over the 95\% of them.
%In addiction, these apps doesn't contain in-app advertisements, regardless they are free.
%%%%%%%%%%%%%%%%%%%%%%%%%%%%%%%%%%%%%%%%%
%%% packet's protocol on captures
%%%%%%%%%%%%%%%%%%%%%%%%%%%%%%%%%%%%%%%%%
% 
% protocol presence in considered apps:
% protocol= TLSv1 54038 
% protocol= HTTP 644
% protocol= TCP 1908
% protocol= SSLv2 63
%total=56653
%%%%%%%%%%%%%%%%%%%%%%%%%%%%%%%%%%%%%%%%%
% app= twitter
% 	 protocol= TLSv1 13435	96.3%
% 	 protocol= TCP 522	3.7%
%	 total = 13957
%%%%%%%%%%%%%%%%%%%%%%%%%%%%%%%%%%%%%%%%%
% app= facebook
% 	 protocol= TLSv1 33398	96.0%
% 	 protocol= TCP 1316	3.7%
% 	 protocol= SSLv2 63	0.1%
%	total = 34777
%%%%%%%%%%%%%%%%%%%%%%%%%%%%%%%%%%%%%%%%%
% app= gmail
% 	 protocol= TLSv1 7205	90.9%
% 	 protocol= HTTP 644	08.1%
% 	 protocol= TCP 70	00.8%
% 	total = 7919

%%%%%%%%%%%%%%%%%%%%%%%%%%%%%%%%%%%%%%%%%%%%%%%%%%%%%%
\section{Machine Learning and Data Mining Background}
\label{ML and DM tools}
In this section, we briefly recall several machine learning and data mining concepts that we use in our paper, while we point the reader to appropriate references for a complete introduction on those topics.

%%%%%%%%%%%%%%%%%%%%%%%%%%%
\subsection{Dynamic Time Warping}
\label{DTW}
Dynamic Time Warping (DTW)~\cite{Muller:2007:IRM:1324818} is a useful method to find
alignments between two time-dependent sequences (also referred as time series) which 
may vary in time or speed. This method is also used to measure the distance or 
similarity between time series.

Let us consider two sequences that 
represent two discrete signals:
$X=(x_1,\ldots,x_N)$ of length 
$N\in\mathbb{N}$; and $Y=(y_1,\ldots,y_m)$  of length $M\in\mathbb{N}$.
DTW uses a local distance measure 
$c:\mathbb{R}\times\mathbb{R} \rightarrow\mathbb{R}_{\geq 0}$ to calculate 
a cost matrix $C\in\mathbb{R}^{N\times M}$, s.t., each cell $C_{i,j}$ reports the distance between $x_i$ and $y_j$. 
The goal is to find an alignment between $X$ and $Y$ having minimal 
overall distance. Intuitively, such an optimal alignment runs along a ``valley'' of 
low cost cells within the cost
matrix $C$. More formally, a \emph{warping path} is defined as a sequence 
$p=(p_1,\ldots,p_L)$ with $p_l=(n_l,m_l)\in[1:N]\times[1:M]$, $l\in[1:L]$ 
satisfying the following three conditions:
\begin{enumerate}
 \item Boundary condition: $p_1=(1,1)$ and $p_L=(N,M)$;
 \item Monotonicity condition: $n_1\leq n_2 \leq \ldots \leq n_M$ and $m_1\leq 
m_2 \leq \ldots \leq m_L$;
 \item Step size condition: $p_{l+1}-p_{l}=\{(0,1),(1,0),(1,1)\}$ for 
$l\in[1:L-1]$.
\end{enumerate}
The total cost of a 
warping path is calculated as the sum of all the local distances of its 
elements. An \emph{optimal warping path} is a warping path $p^*$ 
having 
minimal total cost among all possible working paths. The total cost of an 
\emph{optimal warping path} is also used as a 
distance measure between two sequences $X$ and $Y$. In this paper, we will 
indicate the cost of an \emph{optimal warping path} with $DTW(X,Y)$.

%%%%%%%%%%%%%%%%%%%%%%%%%%%%%%%%%%%%%%%%
\subsection{Hierarchical Clustering}
\label{Hierarchical clustering}  

Hierarchical clustering is a cluster analysis method which 
seeks to build a hierarchy of clusters. This clustering method has the distinct 
advantage that any valid measure of distance can be used. In fact, 
the observations themselves are not required: all that is used is a matrix of 
distances.

In the following we will use a type of hierarchical clustering that is called 
agglomerative: each observation starts in its own cluster, and pairs 
of clusters are merged as one moves up the hierarchy. In order to decide which 
clusters should be combined, a metric (a measure of distance between 
pairs of observations) and a linkage criterion are required. Since we will 
clusterize time-dependent sequences, we will use the total cost of an 
\emph{optimal warping path} as distance metric. As for the linkage 
criterion, that determines the distance between sets of observations as a 
function of the pairwise distances between observations, we will use the 
average 
distance, that is defined as:
$$d(u,v)=\sum_{\substack{1\leq i\leq n\\1\leq j\leq m}}\frac{d(u[i],v[j])}{\left\vert u \right\vert * \left\vert v \right\vert},$$ 
where $d()$ is a distance function, and $u$ and $v$ are two clusters of $n$ and $m$ elements, respectively.
More details about Hierarchical clustering can be found in~\cite{hastie09statisticallearning}.

%%%%%%%%%%%%%%%%%%%%%%%%%%%%%%%%
\subsection{Supervised Learning}
 \label{Classification}
%descrizione algoritmi di apprendimento e in particolare Random Forest, 
%partendo dagli alberi di decisione, random feature selection (necessario dire
% che abbiamo usato anche naive bayes gauss e bernoulli per testare i classificatori?)
%Analysis of a random forests model \cite{Biau:2012:ARF:2503308.2343682}
% Supervised learning is also called classification or inductive 
% learning in machine learning. This type of learning is analogous to human 
% learning from past experiences to gain new knowledge in order to improve the 
% ability to perform real-world tasks. 
Supervised machine learning algorithms learns from labeled instances or 
examples, which are collected in the past and represent past experiences in 
some real-world applications. They produce an inferred model, which can be then 
used for mapping or classifying new instances. An optimal scenario will allow 
for the algorithm to correctly determine the class labels for unseen instances.

In this paper, we will use an ensemble classifier that is called Random 
Forest~\cite{Breiman:2001:RF:570181.570182}.
The main principle behind ensemble methods is that a group of ``weak learners'' 
can be combined together to form a ``strong learner''. 
Random forest leverages a standard machine learning technique called 
 ``decision tree'', which, in ensemble terms, corresponds to the weak learner. 
In practice, it combines together the results of several decision trees trained 
with different portions of the training dataset and different subsets of 
features. More details about the Random Forest classifier can be 
found in~\cite{Breiman:2001:RF:570181.570182}.

\section{Our Framework}
\label{OurFramework}
% review
In this section we describe our framework. In particular, Section~\ref{sec:networkTrafficModellingAndPreprocessing} introduces the 
pre-processing steps that allow us to model the network traffic. Section~\ref{ClassificationUserActions}
describes the methodology used to build training and test dataset, and the procedure used to classify user actions. 

% Then we discuss the methodology for user actions classification, that is reported in Section~\ref{ClassificationUserActions}.
%\subsection{Overview}
%flows Construction -> Clustering of network flows -> Classification of actions

% IP 
% addresses are used to univocally identify the app (or a set of 
% apps) that is generating the traffic. 
% 
% In the following we will detail the various phases that lead to the 
% classification and detection of the actions

% flows construction
% .flow strucutures
% .Domain filtering
% .Packets filtering
% Clustering of network flows
% .Action time interval
% .flow typology identification
% .Distance matrix build method
% .Hierarchical agglomerative clustering
% .Cluster leaders finding
% .flow placing on cluster

%%%%%%%%%%%%%%%%%%%%%%%%%%%%%%%%
\subsection{Network Traffic Pre-Processing}
\label{sec:networkTrafficModellingAndPreprocessing}
%Question once defined an entity we could refere to it with a capital letter? 
%(flow,Packets,etc..)
Mobile apps generally rely on SSL/TLS to securely communicate with 
peers. These protocols are built on the top of the TCP/IP suite. 
The TCP layer receives encrypted data from the above layer, it divides data 
into chunks if the packets exceeds a give size. Then, for each chunk it adds a TCP header
creating a TCP segment. Each TCP segment is encapsulated into an Internet Protocol (IP) datagram, and exchanged with 
peers.
Since TCP packets do not include a session identifier, both endpoints 
identify a TCP session using the client's IP address and the port number.
%Controllare se usare TCP packets o TCP segments
% A packet is a sequence of octets (bytes) and consists of 
% a header followed by a 
% body. The header describes the packet's source, destination and control 
% information. The body contains the data IP is transmitting
% 
% Network traffic on transport layer is composed by packets, that are atomic 
% entities which are characterized by a tuple of fields: \{time, source ip, 
% source port, destination ip, destination port, protocol, length, info \}.
% The value in a Time field is a timestamp given to the packet in the moment of send.
% Source ip and port are rispectly the ip address unresolved and the port of packet's source, same for destination ip and port.
% The protocol value coincide with the transmission protocol used to send the packet, and info field contains additional informations related to the protocol or about errors occurred on transmission.
% 
% In our analysis we doesn't consider encrypted payload of a packet, because SSL 
% encryption is used by many apps (even those ones we considered), so 
% extract significative informations about the content of a packet could be very 
% difficult and out of our target.
% As reported in \cite{Lim:2010:ITC:1921168.1921180} the most important features on
% traffic classification are those related to the ports used for comunication and 
% the size of packets exchanged.

\begin{figure} [h!]
       \centering
    \epsfig{file=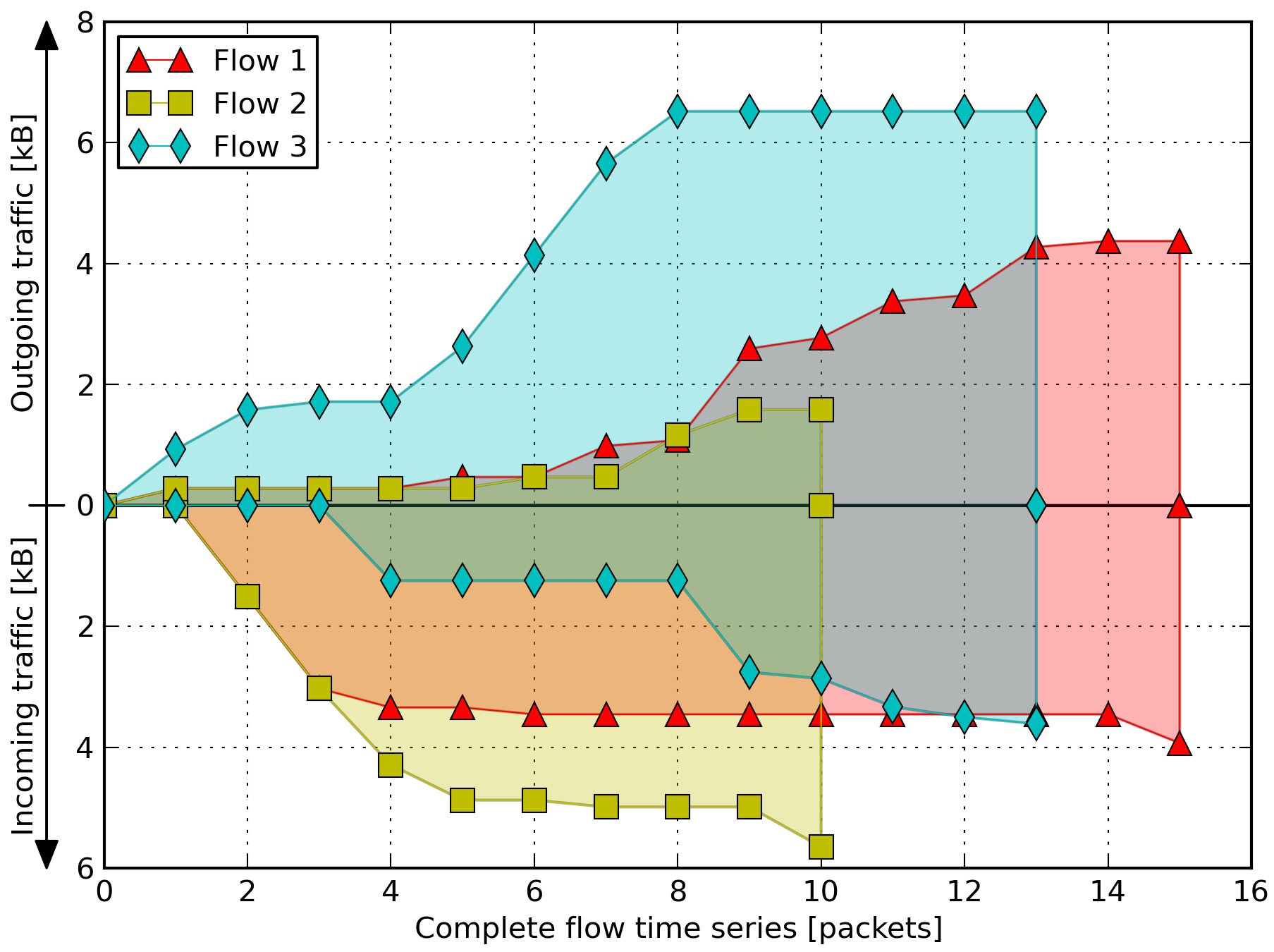,width=8.5cm}
    \caption{Representation of flows time series.}
    \label{fig:fbflowsoverlap}  
\end{figure}

A fundamental entity considered in this paper is the traffic \emph{flow}: with this 
term we indicate a time ordered sequence of TCP packets exchanged between two peers 
during a single TCP session. 
We model each network flow as a set of time series:
% complete time series, incoming time series, outgoing time series
(i) a time series is obtained by considering the bytes transported by 
incoming packets only; (ii) another one is obtained by considering bytes transported by 
outgoing packets only; (iii) a third one is obtained by combining (ordered by time) bytes transported by both incoming and 
outgoing packets. Additional time series may be obtained for example
by considering other parameters such as the time-gap between different packets. %Furthermore, other, or the number of packets per burst in each direction. 
However, we will only use the first three types mentioned above.
\hilight{The use of time series allows to represent network flows making them independent from the properties of the connection. 
%For examples, this allows to compare flows from networks with different bandwidth or signal quality.
}
%\hilight{Time series provide a representation of a flow that does not depend from the connection features. Indeed, using this representation we can compare flows %from networks with different bandwidth and signal quality, so not affected by delays or errors}  
% Additional 
% time series are obtained from those above mentioned by removing a given number of 
% packets at the beginning or at the end of the flow. 
% This simple mechanism allows us to focus on portions of the flow that may be 
% more important than others. Indeed, experimentally we find out that for certain 
% apps the first part of the flow is more significant that the last part, 
% while for other apps it holds the opposit.
Table~\ref{tab:exampleOfTS} reports an example of time series generated from three network flows, while Figure~\ref{fig:fbflowsoverlap}
graphically represents these flows through a cumulative chart. The lower side of the chart represents incoming traffic, while
the upper side represents outgoing traffic. This is only one of the possible representations, and it shows that the ``shapes''
of these three network flows are quite different. Intuitively, our classification approach aims to identify the ``shape'' of \hilight{unseen flows}.

\begin{table*}[ht!]
       \centering
        \begin{center}
\begin{tabular}[t]{|l|l|l|}\hline
\textbf{Flow ID} & \textbf{Time series type} & \textbf{Time series}\\\hline
\multirow{2}{*}{Flow 1} & Incoming & [1514, 1514, 315, 113, 477]\\
 & Outgoing & [282, 188, 514, 96, 1514, 179, 603, 98, 801, 98]\\
 & Complete & [282, -1514, -1514, -315, 188, -113, 514, 96, 1514, 179, 603, 98, 801, 98, -477]\\\hline
 \multirow{2}{*}{Flow 2} & Incoming & [1514, 1514, 1266, 582, 113, 661]
\\
 & Outgoing & [282, 188, 692, 423]\\
 & Complete & [282, -1514, -1514, -1266, -582, 188, -113, 692, 423, -661]
\\\hline
  \multirow{2}{*}{Flow 3} & Incoming & [1245, 1514, 107, 465, 172, 111]
\\
 & Outgoing & [926, 655, 136, 913, 1514, 1514, 863]
\\
 & Complete & [926, 655, 136, -1245, 913, 1514, 1514, 863, -1514, -107, -465, -172, -111]
\\\hline
        \end{tabular}
        \end{center}
\caption{Example of time series generated from three network flows. Values within square brackets
represent the amount of bytes exchanged per packet: negative values in complete time series indicate incoming bytes,
while positive values indicate outgoing bytes.}
\label{tab:exampleOfTS}
\end{table*}

%TODOspecificate che i flussi considerati vengono generati dalla stessa azione

Before generating for each flow the corresponding set of time series, 
a few pre-processing steps have to be 
performed. In particular: 1) we apply a domain filtering to select only flows belonging to the analyzed 
app; 2) we filter the remaining flows, in order to delete packets that 
may degrade the precision of our approach (i.e., we filter out ACK and retransmitted 
packets); 3) we limit the length of the generated time-series.
In the following, we will detail these three pre-processing steps.

%%%%%%%
\paragraph{Domain filtering}
\label{Domain filtering}

\hilight{The network traffic generated by an application is generally directed toward a back-end infrastructure, that may be identified with a single server, or a set of servers that may be even behind a load balance. Since we analyze each app independently, we need to make 
sure that traffic generated from apps other than the considered one (or traffic generated by the OS) do not interfere with the 
analysis. Different methods can be used in order to identify the app that generated each network flows. The destination IP address is a trivial discriminating parameter. However, in case of a load balanced back-end, we should know all the individual IP addresses that can be involved in the communication. The same happens when the back-end is composed by several components such as different web services, databases, etc. To overcome this problem we use another strategy: we take into consideration for further analysis only the flows which destination IP addresses owners have been clearly identified as related to the considered app. In the implementation of our framework, we leverage the WHOIS protocol for this purpose, but we want to highlight that this is only one of the possible way. Business and other context information may be used in order to perform the domain filtering.
% Information about an IP address that is provided by a reverse-DNS look up may not be sufficient to determine the owner of that IP address.
% For this reason, we use WHOIS protocol to perform a series of look ups, inferring
% more information about the owner of the destination IP address (e.g., name and address of the registering organization).
% We take into consideration for further analysis only the flows which destination IP addresses owners have been clearly identified as related to the considered app.
}
%When the owner of the IP address is clearly related to the 
%app which is subject of the analysis, we take the flow into consideration.}% for further analysis.} 
We also take into consideration the traffic related to third parties services (such as Akamai or Amazon) that are indeed used by several applications~\cite{Wei:2012:PMP:2348543.2348563}. 
% \hilight{We underline that the apps considered in our analysis own a proprietary back-end to provide their services. There could be the possibility that other apps could offer their services sharing the same domain (e.g., Amazon Elastic Compute Cloud). In that case, we will need another method to insulate the app traffic.
% Domain filtering could be done to filter packets for both classification and recognition phase. This way it will enable to focus the analysis only to a specific app.}

%Potrebbe essere necessario indicare per ogni applicazione analizzata 
%quali servizi terzi sono stati considerati nei flussi raccolti
% For this reason, we consider only flows  resolve each IP 
% 
% leverage on the IP addresses of the destination
% 
% The first step of our method to filter undesire traffic involve the knoledge 
%of ip addresses of packets (expecially for other ip address in a flow).
% Bigger companies reserve for them a specific range of ip addresses.
% With WHOIS command applied on an unresolved ip address performe a look up and 
%retrieve information about owner of that ip address.
% Given the app considered, we collect only ip addresses which 
%informations contains terms related to app (for example terms 
%``\emph{facebook}'',``\emph{akamai}'' and ``\emph{amazon}'' for Facebook app)  

%%%
\paragraph{Packets filtering}
\label{PacketFiltering}

Due to network congestion, traffic load balancing, or other unpredictable 
network behavior, IP packets can be lost, duplicated, or delivered out of order. 
TCP detects these problems, hence requesting retransmission of lost data, and 
reordering out-of-order data. It comes out that several TCP packets that do not 
carry data, may hinder the analysis process.
In the data exchange phase, for example, the receiver sends a packet with the ACK flag set to notify the correct reception of a chunk of data. 
These ACK packets are transmitted in asynchronous mode so they are affected by many factors related to round trip time of the connection link. 
The order of the received packets may hinder the evaluation of the similarity between two network flows.
For this reason, we filter out all packets retransmissions, as well as packets marked with the ACK flag.
\hilight{Note that the metric that we will use in order to measure similarity between flows (see Section }
\ref{ClassificationUserActions}) \hilight{will mitigate the consequences of a missing packet.
% 
% A packet is retransmitted in two cases: when its content contains an error or in case of not received. 
% In the first case, the size of packet with error could vary, while in the second a packet results missing.
% Fortunately, DTW metric consider a the whole flow, mitigating the error in packets sequences alignment in both of those cases
}
We filter out also other packets that do not bring any additional information helpful in characterize flows. %TODO review
%For example, 
In particular, we filter out the three way handshake executed to open a TCP connection, and the packets exchanged to close it.

\paragraph{Timeout and packets interval}
\label{ActiveInactiveTimeouts}

%flow limited 
% flows usually contain different number of packets, even if generated by the 
%same interaction, many of them are related to the current status of 
%app.
% quanto sono settati negli esperimenti? Per ora ho commentato il paragrafo 
% sottostante
% A part in a communication is considered as a couple (ip 
% address, port).
% Packets regrupped in the same flow share the same two parts involved in the 
% communication, where one part is considered as reference (usually ip address of 
% our devices). 
% a flow can be considered as a tuple \{ reference ip, reference port, other ip, 
% other port, packet sequence, start time, end time\}, where ``\emph{start time}'' and ``\emph{end 
% time}'' are respectly the timestamp in the first packet and the last one.
% Our algorithm constructs flows one packet at the time and so it requires the 
% whole set of packets in a capture, ascendly ordered by value in time field, and 
% two temporal contraint for flow termination: they are active and inactive 
% timeouts parameters (both in terms of seconds).

Two different techniques are used to limit the length of the generated time series: 
a \emph{timeout} mechanism and the specification of a \emph{packets interval}. 
The \emph{timeout} mechanism is used to terminate the flows that did not receive any new packet
since $4.5$ seconds. Indeed, it has been proved experimentally that 95\% of all packets arrive at most $4.43$ seconds after
their predecessors~\cite{Stober:2013:YSY:2462096.2462099}. The \emph{packets interval} 
specifies the first and the last packet to be considered. 
% which packets we want to consider. 
% limit the length of the time series obtained from a 
% flow by setting an interval that specify the first and the last packet to be considered. 
% On the contrary, the latter is applyed on the time series in order to limit 
% their size in terms of 
% 
% 
% 
% Flow related to an action performed by the user on a device are likely to happen in a ``short" 
% time interval after the action itself. We therefore consider an active timeout to limit the 
% the total flow duration. 
% % In order to identify this time interval, we consider two timeouts: \emph{active} and \emph{inactive} timeout.
% % The active timeout limits the total flow duration. 
% In particular, if the total duration 
% of the flow exceeds $4.5$ seconds, the current flow is terminated and a new one is started. 
% 
% 
% On the 
% other hand, inactive timeout is used to terminate a flow that represents 
% an open TCP connection that did not receive any new packet for a while (as specified by the timeout). 
% To cite practical numbers for those parameters, in 
% the experimental section we fixed the active timeout to $10$ seconds, and the 
% inactive timeout to $4.5$ seconds~\cite{Stober:2013:YSY:2462096.2462099}. Furthermore, in our solution we limit the length of the time series obtained from a 
% flow by setting an interval that specify the first and the last packet to be considered. 
For example, considering a flow $f$ composed by $l$ packets, and the interval $\left[x,y\right]$ with $x\leq y$ and $y\leq l$, 
the corresponding time series will be composed by $y-x+1$ values that report the bytes of the 
$x^{th}$ to the $y^{th}$ packet. 
This simple 
mechanism allows us to focus on particular portions of the flow. The first part, for example, is often the 
more significant. In the experimental part, we report the results for different 
configurations of packets intervals, showing that the best configuration is 
app dependent.  

\subsection{Classification of User Action} 
\label{ClassificationUserActions}
%  Detailed description of the clustering technique that we use. 
  
%\subsubsection{Creation of a Labeled Dataset}
\label{LabeledDataset}
Since we use a supervised learning approach, it is necessary to create a labeled dataset that
describes the user actions that we want to classify. 
In order to build this dataset, we simulate a series of user actions, and for each one we identify the 
flows generated after the execution of the action itself. For each app that we analyze
we focus on actions that are significant for that particular app. 

%\label{FlowsCategories}
% A user 
% action may involve a single interaction (tap or swipe on the touchscreen) or 
% multiple interactions (sequence of keys pressed on keyboard). 
In most cases, a single user action generates a set of different flows (i.e., not just a single one). 
Furthermore, different user actions may generate different sets of flows. 
Our classification method is based on the detection of the sets of flows that 
are distinctive of a particular user action.
In order to elicit these distinctive sets of flows, we build clusters
of flows by using the agglomerative clustering approach described in 
Section~\ref{Hierarchical clustering}. 
\hilight{Flows that are similar one to each other will be grouped together in the same cluster, while not similar flows 
will be assigned to in different clusters.} The average distance is used as 
linkage criterion, while the computation of the distance between two flows combines
the distances of the corresponding time series. 
% Indeed, since a single flow is represented as a set of time series, we 
% use a function to combine the distances between two flows $f_i$ and $f_j$. 
Supposing that each 
flow $f_i$ is decomposed into a set of $n$ time series 
$\{T_1^i,\ldots,T_n^i\}$, the distance between $f_i$ and $f_j$ is defined as:
$$dist(f_i,f_j)=\sum_{k=1}^{n} w_k\times DTW(T_k^i,T_k^j),$$
where $w_k$ is a weight assigned to the particular time series. Weights can be 
assigned in such a way to give more importance to some type of time series with 
respect to others. For example, it is possible to give more weight to the 
time series that represent incoming packets, and less weight to those that 
represent outgoing packets. %TODO CHECK

In order to reduce the computational burden of the subsequent classification, a 
leader is elected for each cluster. 
Leaders will be the representative flows of their clusters. 
%, and they will be used both during the training and the test phase. 
Given a cluster $C$ containing the flows 
$\{f_1,\ldots,f_n\}$, the leader is elected by selecting the flow $f_i$ that has 
minimum overall distance from the other members of the cluster, that is: 
$$\argmin_{f_i \in C}\left( \sum_{j=1}^n dist(f_i,f_j)\right).$$
Clustering is executed over the set of flows that will be used to build the training dataset. 
\hilight{The cluster leaders will be used to build both the training and the test datasets. The user 
actions will be the instances of the datasets, while the class of each instance is a label representing 
the action. We will have one integer feature for 
each cluster identified through the agglomerative clustering. The value of
each feature is determined 
by analyzing the flows related to an action.}
Each flow $f$ captured after the execution of an action will 
be assigned to the cluster that 
minimizes the distance between $f$ and the leader of the cluster.
The $k^{th}$ feature 
will therefore indicate the number of flows that have been 
assigned to the cluster $C_k$ after the execution of that action.
%TODO spiegare meglio
For example, for the 
action \emph{send mail}, the $k^{th}$ feature will be equal to $2$ 
if there are $2$ flows labeled with \emph{send mail}
assigned to the cluster $C_k$.

Finally, we execute the classification with Random Forest algorithm. 
The main idea behind the overall approach is that different actions will ``trigger'' different
sets of clusters. The classification algorithm will therefore learn which are these sets, 
and will be able to correctly determine the class labels for unseen instances. 

%Naturally, these instances may have different attributes set to 
%$1$. 
  %Detailed description of the classification procedure that we use.
  %Clusters produced in the previous phase are considered as features for classification.
  %Given an action time interval $O$ and all the flows inside it, we consider the number
  % of flows those belong to a cluster with index $c$ as the value for the feature related to $c$ for the $O$ action time interval 
  %TODO parliamo qui del classificatore usato? (Random forest classificator con 60 estimatori)

%%%%%%%%%%%%%%%%%%%%%%%%%%%%%%%%%%%%%%%%%%%%%%%%%%%%%%%%%%%%%%%%%%%
\section{Experimental Results}
\label{ExperimentalResults}
%\subsection{Data acquisition}

In order to assess the performance of our proposal, we considered three widespread apps: Gmail, Facebook and Twitter. 
\hilight{We select these apps because of their high popularity}~\cite{androidRank}.
\hilight{Indeed, Gmail is the world's largest email service}~\cite{gmailbeatshotmail},
\hilight{and its Android app has over one billion downloads.}
\hilight{On the other hand, Facebook and Twitter are not only the most popular Online Social Networks}~\cite{ebizmbaRank},
\hilight{but they also had a leading role in the Arab spring}~\cite{fbtwarabstring}
\hilight{and the Istanbul's Taksim Gezi Park protests}~\cite{fbtwturkey2}
\hilight{(when Turkish government blocked Twitter)}.
%\hilight{Moreover, these particular Android apps need Internet connection to work.}
\hilight{We believe that the results we obtain in our analysis also hold for other apps that provide similar functionalities (e.g., Yahoo mail, WhatsApp or LinkedIn), while a thorough evaluation of this claim is left as future work. }%TODO
To collect the network traffic related to different user actions, we set up a \hilight{monitored environment.} 
In this section we present the elements that compose this environment (Section~\ref{HDWandNetworkConfiguration}), 
the methodology used to collect the data (Section~\ref{datacollection}), and the results of the evaluation (Section~\ref{ClassificationPerformances}).

%%%%%%%%%
\subsection{Hardware and Network Configuration}
\label{HDWandNetworkConfiguration}

%In the very first part of this description we have to discuss the data acquisition system 
%by the hardware point of view, describing the principles features of components

%\subsubsection{Devices and Software}
For the evaluation of our solution, we used a Galaxy Nexus (GT-I9250) smartphone, running the Android 4.1.2 (Jelly bean) 
operative system.
% Being a ``Google" device, this device was running an original version of Android OS, without any other proprietary software. 
%MC: non farei sorgere dubbi al reviewer
%(that could interfer with our analysis).
%However the model of the device or other software do not affect the analysis.
% Anyway, we recall that in our approach we discard traffic flows having an IP address with domain not related to a considered app (as explained in Section~\ref{Domain filtering}).
We enabled the ``\emph{Android Debug}'' option in order to allow the usage of the ADB (Android Debug Bridge) interface via USB cable. 
We used a Wi-Fi access point 
(U.S. Robotics USR808054) to provide wireless connectivity to the mobile phone. 
Finally, we used a server (Intel Pentium Processor dual core E5400 2.7GHz with 4 GB DDR2 RAM)
with two network cards running Ubuntu Server 11.04 LTS to route the traffic from the access point to the Internet, and vice versa. 

%\subsubsection{Access to network}
% Hardware components we needed for let devices be able to access to Internet are
% a wireless access point and a PC with two ethernet network cards.
% Starting from the devices side, we use a wireless access point (U. S. Robotics 
% USR808054) to provide them wireless connectivity. The access point was connected
% to first network card via an Ethernet cable, PC was configured in order to use this
% network card as gateway. Then from the other network card, the PC was connected to 
% main Department network access to Internet. The data acquisition software and scripts
% runs on a Ubuntu Server 11.04 LTS that makes a bridge between PC’s network card, 
% and allowing Internet connectivity.
%\subsubsection{Wireshark}
To eavesdrop network packets flowing through the server, we used Wireshark 
software. From a Wireshark capture file, we 
created a comma separated file (csv), where each row describes a
packet captured from the access point's interface. 
For every packet we reported source and destination IP addresses, ports, size in bytes and
time in seconds from Unix epoch (i.e., 00:00:00 UTC, 01 January 1970), protocol type and TCP/IP flags. %packet info 
Since the payload is not relevant to our analysis, it has been omitted. 
This data have been then used to generate the time series as 
explained in Section~\ref{sec:networkTrafficModellingAndPreprocessing}.

%\newpage
%%%%%%%%
\subsection{Dataset Collection and Analysis}
\label{datacollection}

For our study we considered three apps installed from the official Android market: Gmail v4.7.2 , Facebook v3.8, and Twitter v4.1.10.
For each app, we created 10 accounts that have been %TODO va bene 10 accounts?
considered in two different categories of users: ``\emph{active}'' and 
``\emph{passive}'' users. ``\emph{Active}'' users simulated the behavior of 
users that actively use the app by sending posts, email, tweets, surfing 
the various menus, etc. ``\emph{Passive}'' users simulated the behavior of 
users that passively use the app, just by receiving messages or posts. 
The accounts of both passive and active users have been configured in such a way 
to have several friends/followers within the group. We avoided to 
configure the accounts with actual friends or followers, in order to avoid 
interference
due to notifications of external users activities that were not under our control.

%\subsection{Application scripts}
To reach a particular target, a user may have to perform several actions in 
a precise order. An action could 
be simple (e.g., a tap on a button, a swipe, or a selection of 
edit box), or complex (e.g., type a text, which is a sequence of keyboard inputs).
For example, a user has to perform three actions in a precise sequence to post a message on his Facebook wall.
He has to be sure that the Facebook app shows the ``\emph{user's wall}'', 
then he has to tap on the ``\emph{write a post}'' button (1), %action 1
fill the edit box with some text (2), %action 2
and finally tap on the ``\emph{post}'' button (3). %action 3
% For example, in order to post a message on his Facebook wall, a user has to 
% be sure that the Facebook app shows the ``\emph{user's 
% wall}'', then he has to tap on the ``\emph{write a post}'' button, %action 1
% fill the edit box with some text, %action 2
% and finally tap on the ``\emph{post}'' button. %action 3
%TODO difference between action and operations
It is important to highlight that we do not use static text to fill in 
text boxes, but the text is randomly selected from a large set of sentences.
A script submits the sequence of actions to the
mobile phone through the ADB commands, and it 
captures the network traffic that is generated.
\hilight{The script records also the execution time of each action.
%, and it waits 20 seconds before executing a new action. %maybe Next
By using the recorded execution time of each action, it is then possible to label the flows extracted from the 
network traffic with the user action that produced it.}
% \hilight{At this point, it result clear that simulating user action and collecting their traffic is not so easy.
% It requires a lot of precision and the script for an app must be written manually.
% For this reason in the experiments we report in this paper, we consider a small number of highly relevant apps.}
\hilight{For each app, we choose a set of actions that are more sensitive than others from user privacy point of view (e.g., send an email or a message, for the reasons we report in} Section~\ref{Introduction}).
The list of these actions is reported in Table~\ref{tab:appcations}. 
\hilight{We underline that we do not ignore other user actions, but we label them as \emph{other}.
In such a way we have several benefits}~\cite{mitchell1997machine}:
\hilight{we obtain a greater representation of data in terms of variety and variance of examples; 
we reduce the chances of overfitting; we improve the performance of the classifier on relevant user actions.}
%\hilight{Indeed, \textit{other} actions are useful to}

\hilight{We collected and labeled the traffic generated by 220 sequences of actions for each app, where a sequence is composed by 50 types of actions 
(for a total of $11660$ examples of actions for Gmail, $6600$ for Twitter, and $10120$ for Facebook).
%53 per gmail, 30 per twitter, 46 per facebook
The user action examples in the dataset was divided in a training set  and a test set.
We use the training set to train the classifier, while we use the test set to evaluate its accuracy. 
%The 70\% of these actions have been used for the training set, while the remaining 30\% have been used for the test set. 
We underline that to build the test set we used accounts that have not been used to create the training set.
By using different accounts to generate 
the training and the test set, it is possible to assure that the results 
of the classification do not depend on the specific accounts that have been 
analyzed.}

\begin{table}[h!] 
\begin{center} {%\footnotesize
\begin{tabular}{|l|l|}
\multicolumn{2}{c}{}  \\
  \multicolumn{2}{c}{\textbf{Facebook}}  \\
  \hline
  \textbf{Action} & \textbf{Description}\\
  \hline
  \emph{send message} & {send a direct message to a friend} \\
  \emph{post user status} &  { post a status on the user's wall}\\
  \emph{open user profile} &   {select user profile page from menu}\\
  \emph{open message} &	{select a conversation on messages page}\\
  \emph{status button} & 	{select ``\emph{write a post}'' on user's wall} \\
  \emph{post on wall} & 	{post a message on a friend's wall}  \\
  \emph{open facebook} & 	{open the Facebook app} \\
 
 \hline
 \multicolumn{2}{c}{}  \\
 \multicolumn{2}{c}{\textbf{Gmail}}  \\
 \hline
  \textbf{Action} & \textbf{Description}\\
  \hline
 \emph{send mail} &  { send a new mail} \\
\emph{reply button}    &  {tap on the reply button} \\
  \emph{open chats}    &  {select chats page from menu}\\
\emph{send reply} &   {send a reply to a received mail}\\

  \hline
  \multicolumn{2}{c}{}  \\
  \multicolumn{2}{c}{\textbf{Twitter}}  \\
  \hline
  \textbf{Action} & \textbf{Description}\\
  \hline
  \emph{refresh home}   &  { Refresh the home page}  \\
  \emph{open contacts}   &  {select contacts on menu}   \\
  \emph{tweet/message}&  { publish tweet or send message}\\ 
  \emph{open messages}  &  { select direct messages page}\\
  \emph{open twitter}    &  { open the Twitter app}\\
  \emph{open tweets}     & { select tweets page}\\

  \hline
\end{tabular} }
\end{center}
\caption{Description of the relevant actions for each app.}
\label{tab:appcations}
\end{table}
  %\subsection{Description of the Scenario}
%   In implementation of our framework, in particular on Clustering paramenter selection, we focus on
%   actions with high significance among all actions considered for an 
% app.

\begin{figure}[h!]
    \centering
\epsfig{file=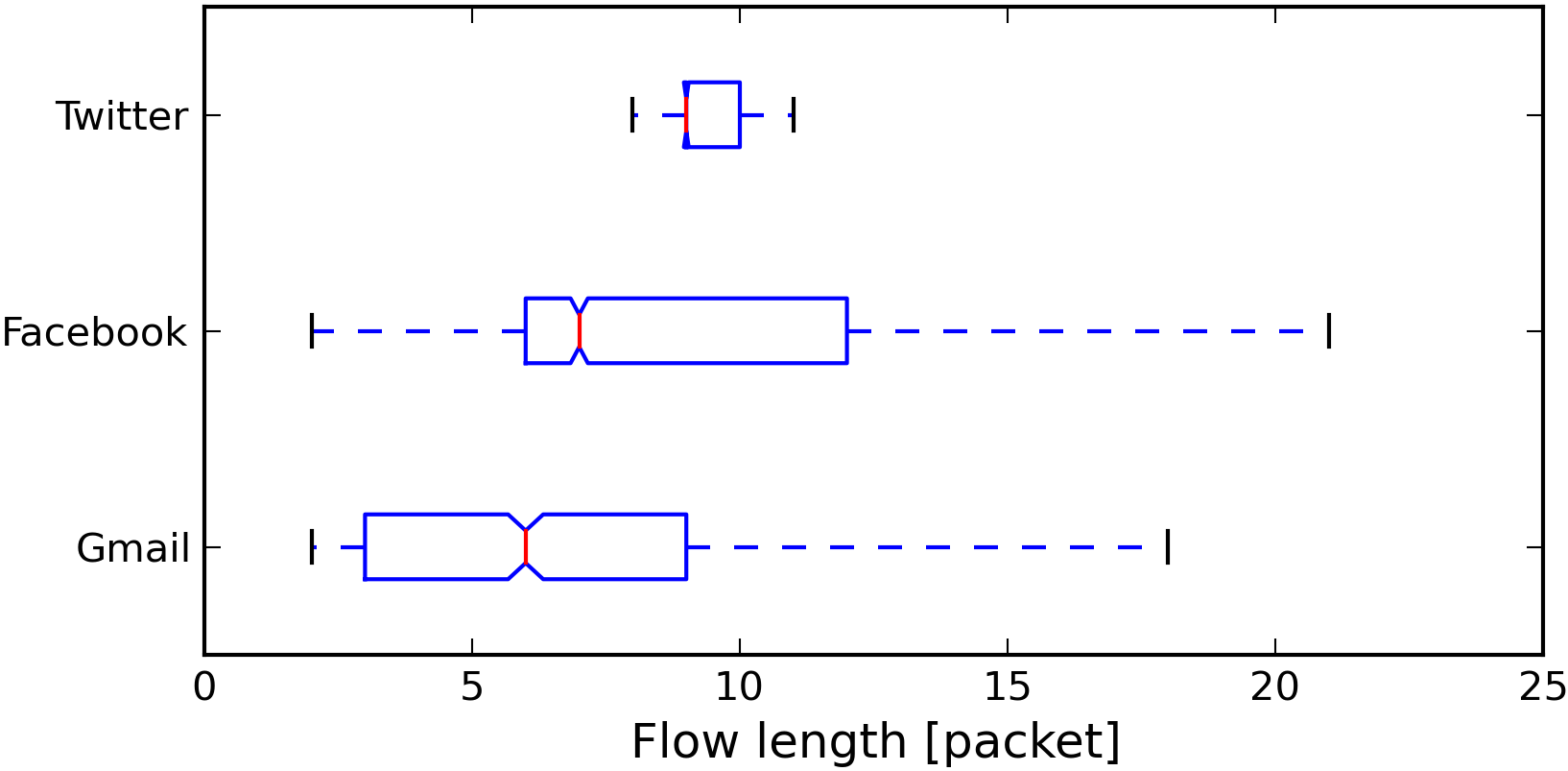,
width=\linewidth}
    \caption{Statistical distribution of the length of the complete time series 
extracted from the network traffic. First and third quartile are 
represented as the left and right side of the notched box. The notch of the box 
represents the median value. Lines that extend horizontally from the boxes 
indicate the $2^{nd}$~percentile (left) and the $98^{th}$~percentile (right).}
    \label{fig:distributionflowlength}
  \end{figure}

%   \subsubsection{Clustering parameters selection}
%   \label{ClusteringParametersSelection}
  %argomentazione su quali sono state le considerazioni che ci hanno portato a scegliere i parametri per il calcolo delle distanze tra flow time-series
  %preclustering all'interno delle singole operazioni
  %grafici strutture twitter/gmail
As explained in Section~\ref{sec:networkTrafficModellingAndPreprocessing}, 
each network flow is modeled as a set of time series. 
% One of them is 
% obtained by considering the sequence of the sizes of the incoming packets, 
% another by considering outgoing packets only, and a third by considering both 
% incoming and outgoing packets. Additional time series are obtained starting 
% from those above by restricting the length of the series. 
Table \ref{tab:weights} reports the weights and the intervals for several configurations (``Conf." in the table) used to limit the length of the time series generated by each app.
We used different weights configurations, and we selected the packets
intervals by analyzing the statistical length of the flows. 
Figure~\ref{fig:distributionflowlength} reports the
statistical distribution of the length of the flows app by app. 
The first quartile, the median and the third quartile are highlighted by 
using a notched box plot. In particular, the median value and the third quartile have been used 
as thresholds to limit the maximum length of the generated time series. 
For the Twitter app, in some cases we set the interval in such a way to focus 
only on the last three or four packets.
Indeed, we noticed that the first part of the time series was identical for each flow.

\begin{table}[h!]
       \centering
        \begin{center}
\begin{tabular}[t]{|l|l|c|c|c|c|}\hline
\textbf{{Apps}} & \textbf{{Sets}} & \textbf{{Weights}} & \textbf{{In}} & \textbf{{Out}} & \textbf{{Complete}} \\
\hline
%%%%%%%%%%% GMAIL %%%%%%%%%%%
\multirow{6}{*}{{Gmail}} 
%%%% short 0 %%%%%%
& \multirow{2}{*}{{Conf.~1}} & 0.80 & [1,4]& [1,2]& [1,6] \\
		                 &  & 0.20 & [1,6]& [1,3]& [1,9]\\ \cline{2-6}
%%%% short 1 %%%%%%
& \multirow{2}{*}{{Conf.~2}} & 0.66 & [1,4]& [1,2]& [1,6] \\
		                 &  & 0.33 & [1,6]& [1,3]& [1,9]\\ \cline{2-6}
%%%% long 1 %%%%%%
& \multirow{2}{*}{{Conf.~3}} & 0.33 & [1,4]& [1,2]& [1,6] \\
		                 &  & 0.66 & [1,6]& [1,3]& [1,9]\\
\hline %\hline
%%%%%%%%%%% FACEBOOK %%%%%%%%%%%%%
\multirow{6}{*}{{Facebook}}
%%%% short 1 %%%%%%
& \multirow{2}{*}{{Conf.~1}} & 0.66 & [1,3]& [1,5]& [1,7] \\
				 &  & 0.33 & [1,6]& [1,7]& [1,12]\\ \cline{2-6}
%%%% Long 1 %%%%%%%
& \multirow{2}{*}{{Conf.~2}} & 0.33 & [1,3]& [1,5]& [1,7] \\
				 &  & 0.66 & [1,6]& [1,7]& [1,12]\\ \cline{2-6}
%%%% Long 0 %%%%%%%
& \multirow{2}{*}{{Conf.~3}} & 0.20 & [1,3]& [1,5]& [1,7] \\
				 &  & 0.80 & [1,6]& [1,7]& [1,12]\\ \cline{2-6}
\hline %\hline
%%%%%%%%%%%%%% TWITTER %%%%%%%%%%%
\multirow{6}{*}{{Twitter}} 
& \multirow{2}{*}{{Conf.~1}} & 0.95 & - & - &  [7,10] \\ % limit on 10 packets, last 4
		      &  &  0.05 & - & -& [1,10]\\ \cline{2-6}
 & \multirow{2}{*}{{Conf.~2}} & 0.95 & - & -& [8,11] \\ % limit on 11 packets, last 4
		      &  &  0.05 & - & -& [1,11]\\ \cline{2-6}
& \multirow{2}{*}{{Conf.~3}} & 0.95 & - & -& [8,10] \\ % limit on 11 packets, last 3
		      &  &  0.05 & - & -& [1,10]\\ 
\hline

\end{tabular}
        \end{center}
\caption{Weights set configurations and packets intervals for Gmail, Facebook and Twitter apps.}
\label{tab:weights}
\end{table}

\begin{figure} [h!]
        \centering
        \begin{minipage}[b]{0.45\textwidth}
                
\includegraphics[width=8.20cm]{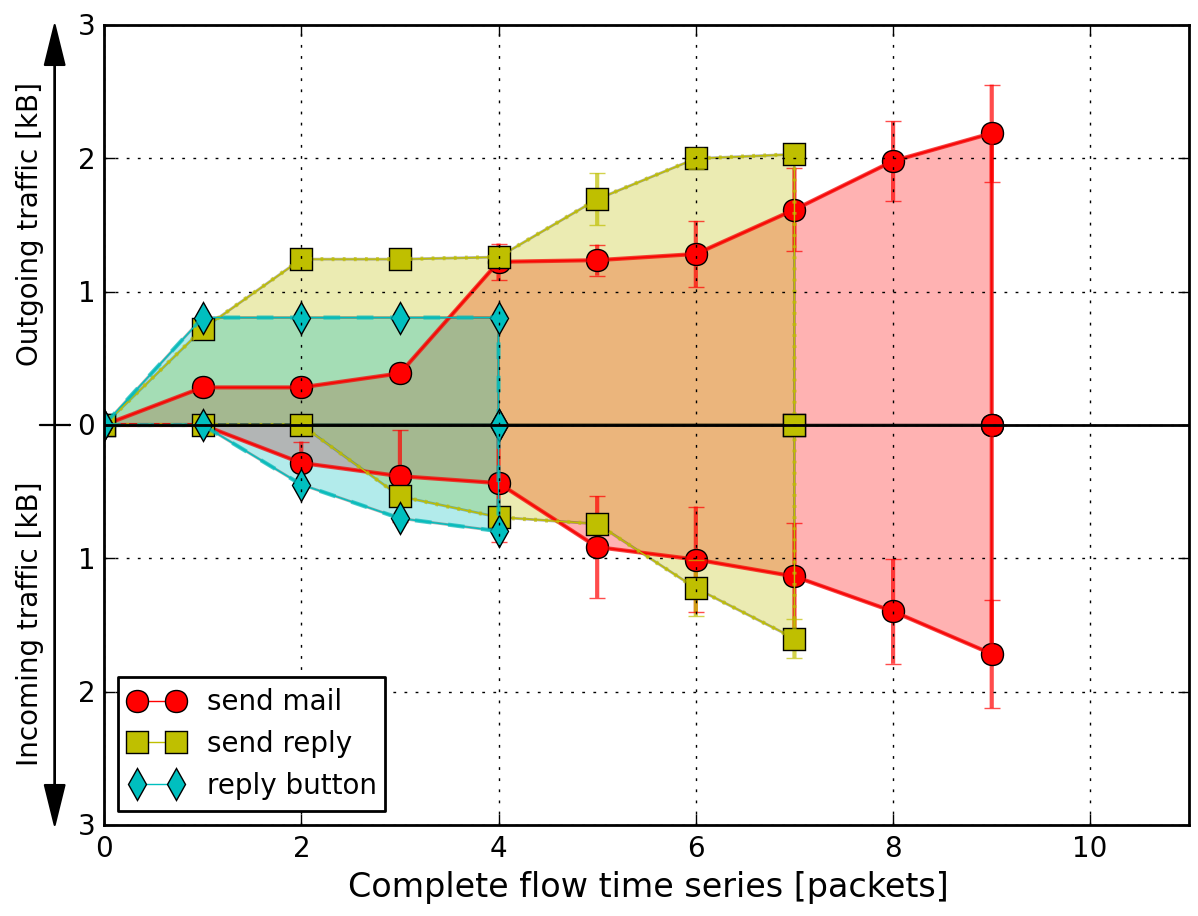}
    \caption{Representation of three different Gmail actions.}
    \label{fig:gmailflows}
        \end{minipage}%
        \\
        \vspace{0.3cm}
         %add desired spacing between images, e. g. ~, \quad, \qquad etc.
        \begin{minipage}[b]{0.45\textwidth}
\includegraphics[width=8.30cm]{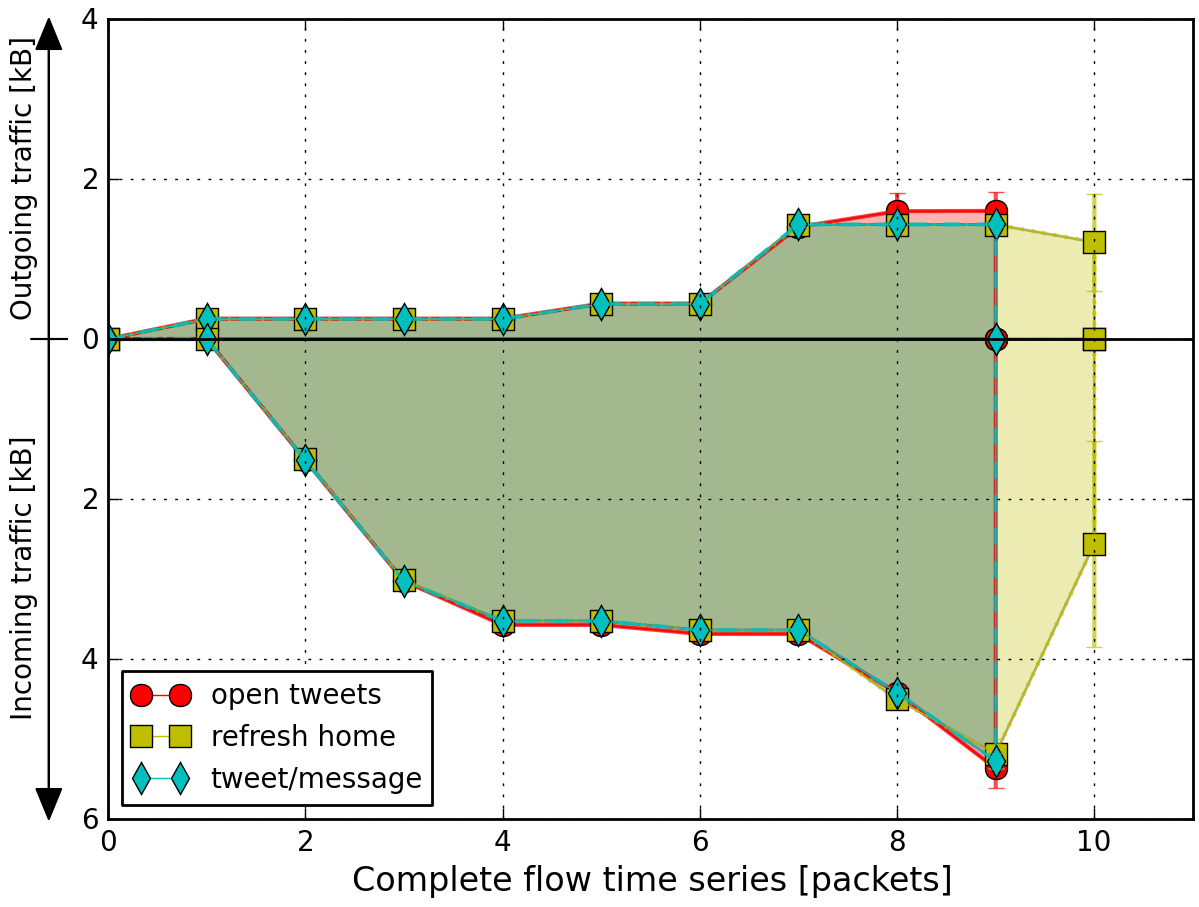}
    \caption{Representation of three different Twitter actions.}
    \label{fig:twitterflows}
        \end{minipage}
       % ~ %add desired spacing between images, e. g. ~, \quad, \qquad etc.
       % \caption{Comparison of three different Gmail and Twitter actions. It can be noticed that Twitter actions are more similar than Gmail actions, indeed  their shapes are largely overlapped.}
       %\label{fig:TwitterGmailComparison}
\end{figure}

% Analyzing Figure 
% \ref{fig:distributionflowlength}, it can be noticed that the the length of the 
% flows generated by Twitter is always between 9 and 10 TCP/IP packets. 
% In fact, 
% this app seems to generate the same type of flows independently of the 
% particular action that is executed. 
To confirm this statement we report in 
Figure \ref{fig:gmailflows} and Figure \ref{fig:twitterflows} the 
graphical representation of the flows that occur when executing three different 
actions in Gmail and Twitter respectively. Comparing the two figures, it can be 
noticed that the shapes of the actions drastically change for Gmail, while they 
are almost unvaried for Twitter. As a matter of fact, different Twitter actions 
just differ in their last packets. Nevertheless, our approach 
reaches very good performance also for this app.
\hilight{In our experiments, we used the Random forest classifier implemented by \textit{scikit-learn} libraries}~\cite{scikitlearn}.
\hilight{The classifier is trained using 40 estimators, with the square root of the number of the features as max feature selection value for each estimator,
and the bootstrap option activated. Each estimator (i.e., weak learner) consists in a decision tree without any restrictions on its depth limit.}

%added temporally
\begin{figure}[h!]
    \centering
    \vspace{0.2cm}
\includegraphics[width=8.60cm]{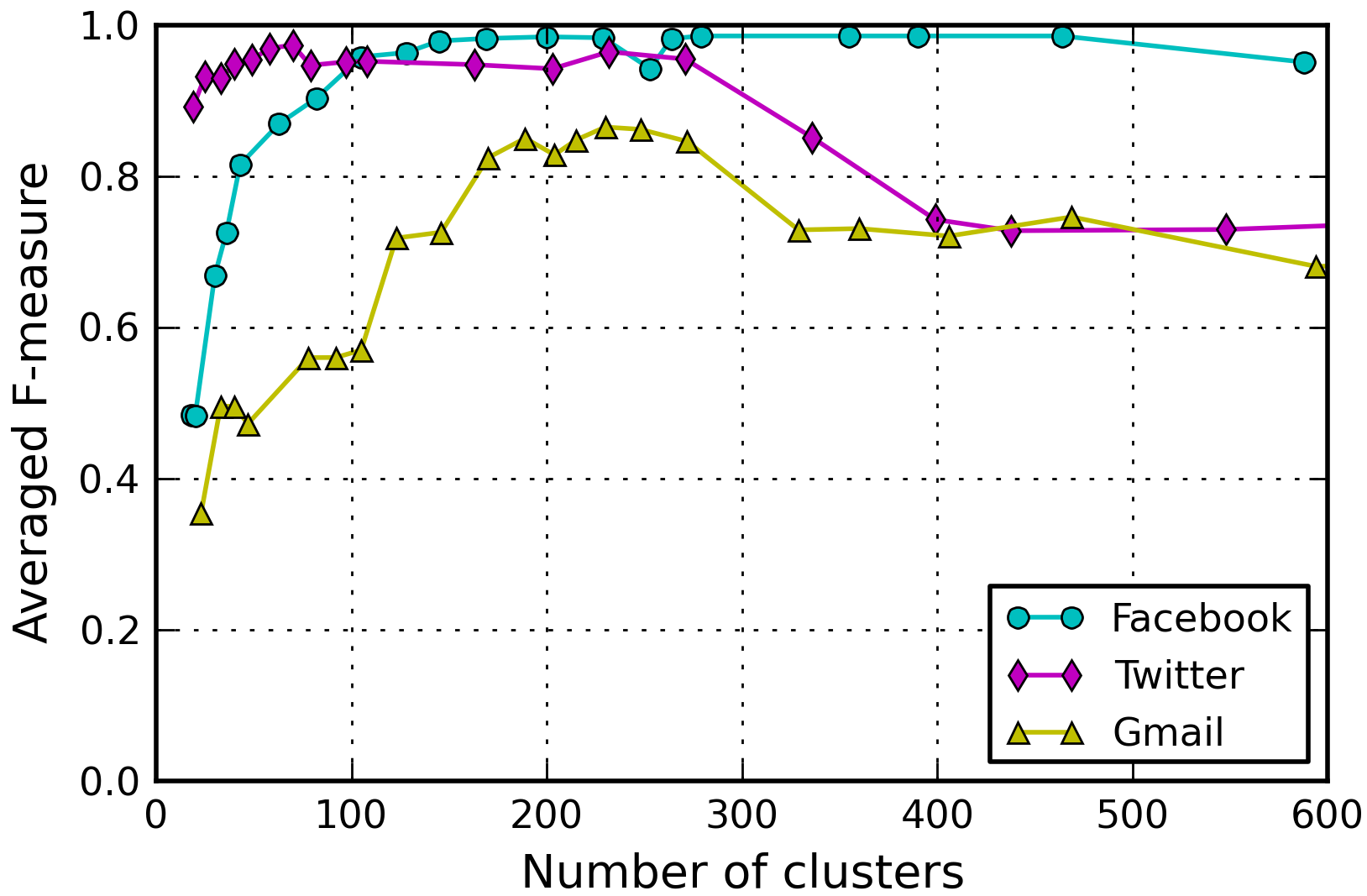}
    \caption{Classification accuracy over number of clusters.}
    \label{fig:accuracyNClusters}
  \end{figure}

  \subsection{Classification Performance}% before called {Results}
  \label{ClassificationPerformances}  

One of the issues to discuss before proceeding to the classification of the user actions
is \hilight{the number of clusters to consider.} %to use in order to group similar flows.} 
In order to establish a reasonable value for this parameter, we used a validation dataset
to study the accuracy of the classification when varying the number of clusters. Figure~\ref{fig:accuracyNClusters}
reports the achieved results. For each app, we therefore considered
the number of clusters that maximized the accuracy, in terms of averaged F-measure. In the following, we report the 
results of the classification app by app. 
In particular, we discuss the average accuracy reached 
when detecting each sensitive user action, we report the corresponding confusion matrices, and 
detailed results for the precision, the recall and the F-measure metrics. %TODO confusion and table for the configuration with best  performances

%%%%%%%%%%%%%%%%%%%%%%%%%%%%%%%%%%%%%  

%%%%%%%%%%%%%%%  
  \subsubsection{Facebook}
  \label{ClassPerf:Facebook}

We focused on seven different actions that may be sensitive when using the 
Facebook app. On average, the F-measure is equal 
to 99\%, with a precision and a recall of 99\% and 98\% respectively. 
Performance reached with different configurations of weights and packets intervals 
constraints are reported in Figure~\ref{fig:facebookperf}. 
For each action at least one of the configurations exceeds 94\% of 
accuracy, while the 
worst performing is always higher than 74\%. %TODO check better
 
Table \ref{tab:facebookreport} reports precision, recall and F-measure reached 
by using {Configuration 3}. We noticed that all the actions have a precision higher 
96\%. The recall is higher than 95\% for all the actions but the \emph{open user profile}, that reaches 91\%. 
%This action corresponds to start of the Facebook app.
In effect we realized that this particular action is 
classified as \emph{other} in 9\% of the examples, %TODO check better
as we can see from the confusion matrix 
reported in Figure~\ref{fig:facebookconfusion}. 

  \begin{table}[h!]
\begin{center} {%\footnotesize
\begin{tabular}{|l|r|r|r|}
  \hline
  %  \multicolumn{5}{c}{Facebook with long0 (Conf. 3)weight set and cut 0.7}  \\
  \textbf{Actions} & \textbf{Precision}& \textbf{Recall} & \textbf{F-measure}\\
  \hline
  \emph{send message} & 1.00 & 1.00 & 1.00 \\ 
  \emph{post user status} & 1.00 & 0.95 & 0.97 \\ 
\emph{open user profile} & 0.96 & 0.91 & 0.94 \\ 
\emph{open message} & 0.98 & 1.00 & 0.99 \\ 
\emph{status button} & 1.00 & 1.00 & 1.00 \\ 
\emph{post on wall} & 1.00 & 0.98 & 0.99 \\ 
\emph{open facebook} & 1.00 & 1.00 & 1.00 \\ 
\emph{other} & 0.99 & 1.00 & 0.99 \\ 
\hline
Average & 0.99 & 0.98 & 0.99 \\ 
  \hline
\end{tabular} }
\end{center}
\caption{ Classification results of Facebook actions by using {Configuration~3}.}

\label{tab:facebookreport}
\end{table}

%%%%%
\subsubsection{Gmail}
\label{ClassPerf:Gmail}
\begin{figure*}
        \centering
        \begin{minipage}[t]{0.45\textwidth}
    \includegraphics[width=8cm]{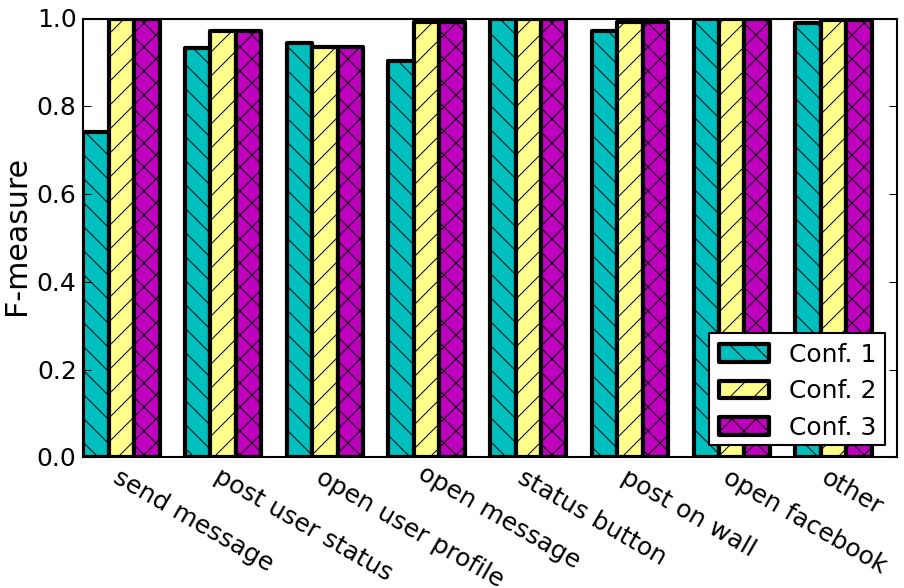}
	\caption{Classification accuracy of the Facebook user actions.}
	\label{fig:facebookperf}
        \end{minipage}%
        \hfill
        \begin{minipage}[t]{0.45\textwidth}
              \centering
\includegraphics[width=8cm]{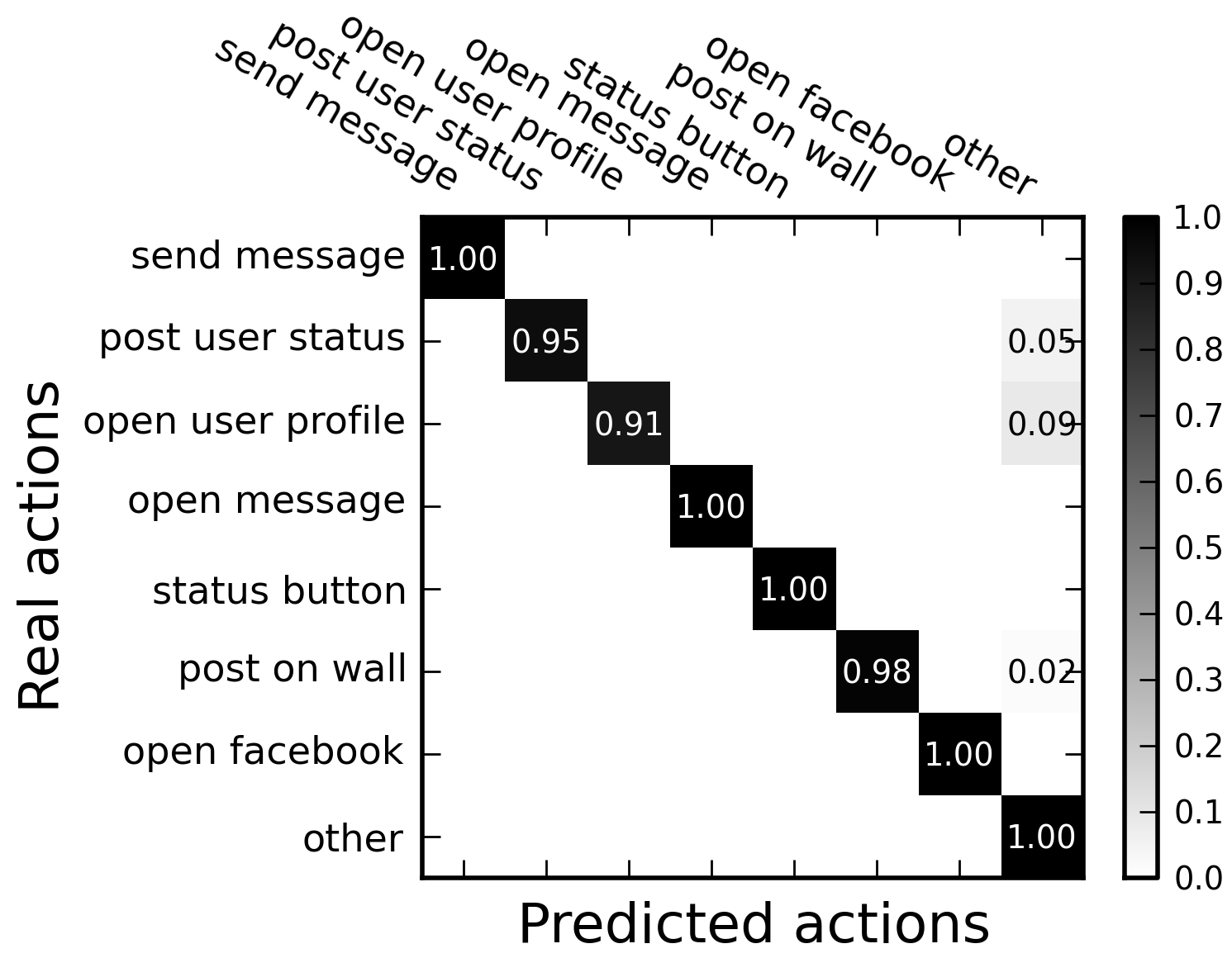}
  \caption{Facebook user actions confusion matrix for Configuration 3.}
  \label{fig:facebookconfusion}  
        \end{minipage}%
        \\
         %add desired spacing between images, e. g. ~, \quad, \qquad etc.
        \begin{minipage}[t]{0.45\textwidth}
	  \centering
	\includegraphics[width=8cm]{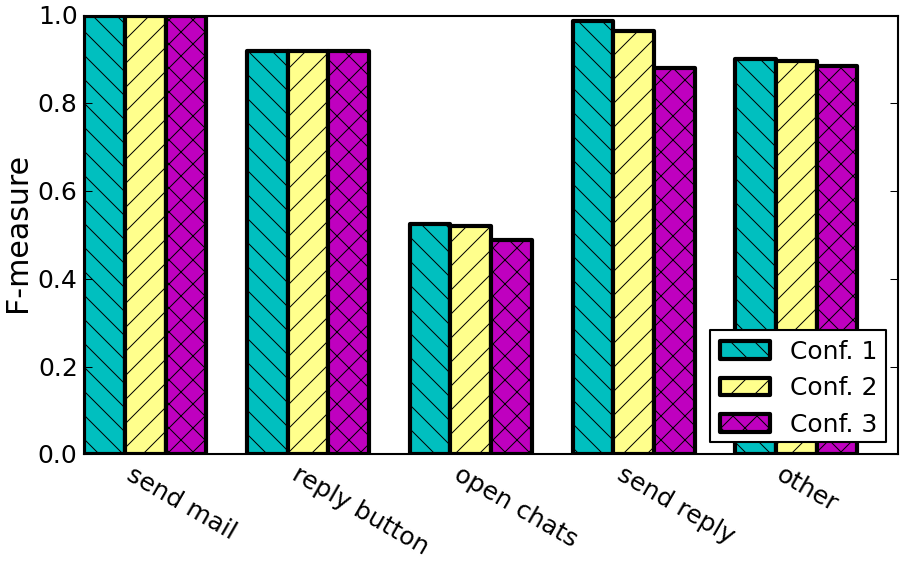}
	  \caption{Classification accuracy of the Gmail user actions.}
	  \label{fig:gmailperf}
        \end{minipage}
        \hfill
        \begin{minipage}[t]{0.45\textwidth}
                \centering
\includegraphics[width=8cm]{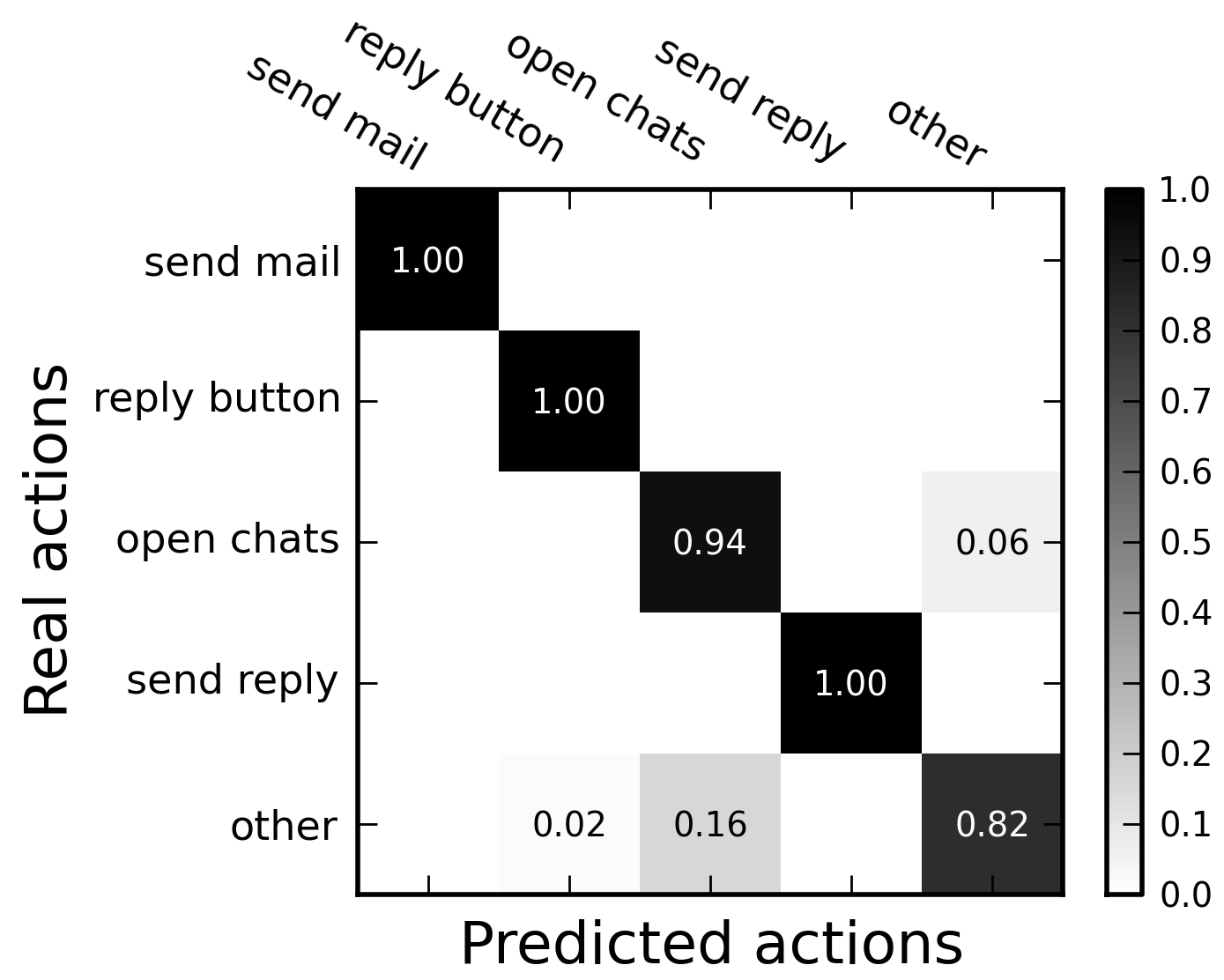}
    \caption{Gmail user actions confusion matrix for Configuration 1.}
    \label{fig:gmailconfusion}    
        \end{minipage}
        \\
        \begin{minipage}[t]{0.45\textwidth}
	  \centering
      \includegraphics[width=8cm]{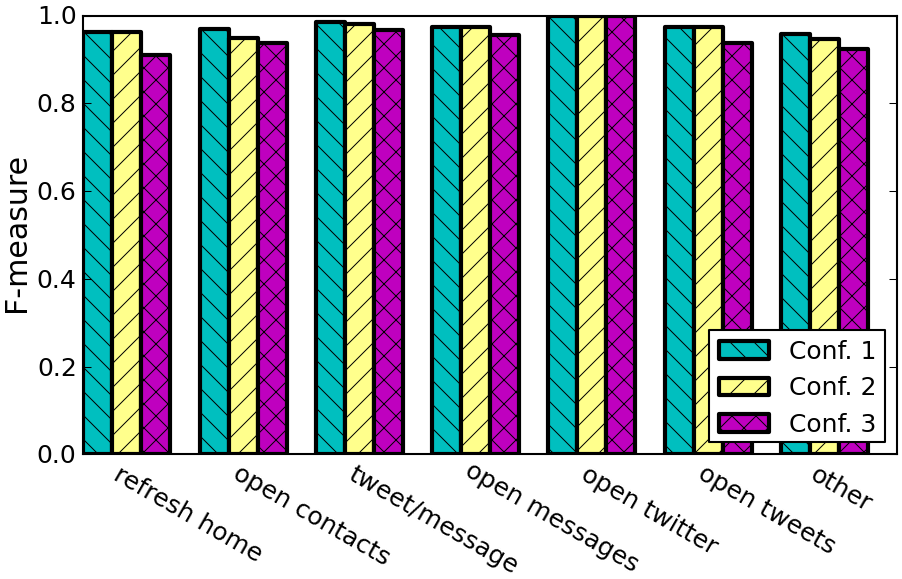}
	  \caption{Classification accuracy of the Twitter user actions.}
	  \label{fig:twitterperf}
        \end{minipage}
	\hfill
	\begin{minipage}[t]{0.45\textwidth}
      \centering
\includegraphics[width=8cm]{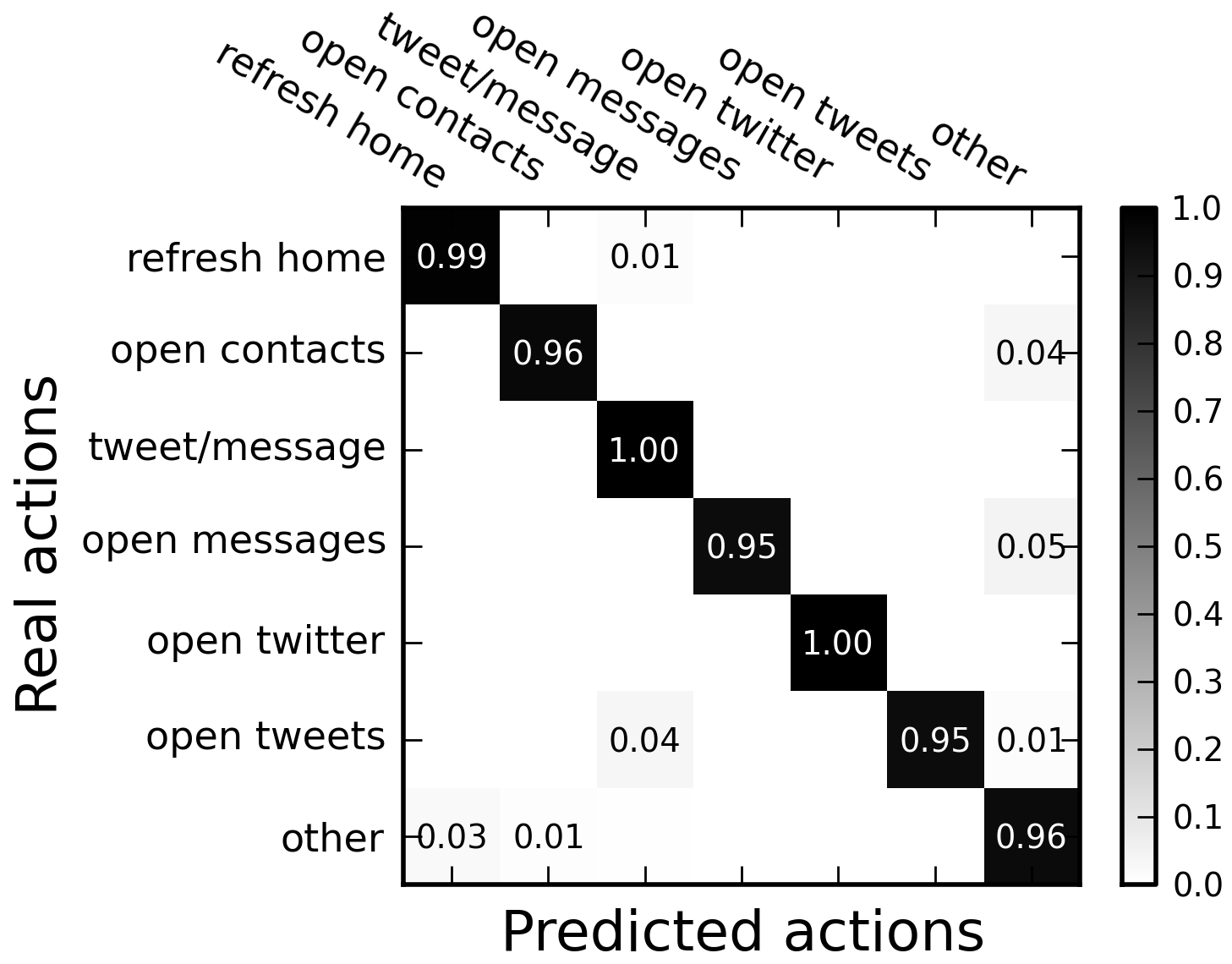}
  \caption{Twitter user actions confusion matrix for Configuration 1.}
  \label{fig:twitterconfusion}           
        \end{minipage}
         %add desired spacing between images, e. g. ~, \quad, \qquad etc.
        %\caption{Classification accuracy and confusion matrix of Facebook, Gmail and Twitter actions.}
  %\label{fig:zzzzz}
\end{figure*}

We analyzed four specific user actions of the Gmail app: \emph{send mail}, 
\emph{reply button}, \emph{open chats} and \emph{send reply}. 
Figure~\ref{fig:gmailperf} shows the classification accuracy that has been 
reached. We observe that we are able to distinguish with high 
accuracy the action of sending of a new mail, from that of replying to a 
previously received message, as well as the tap over the reply button. The 
\emph{open chats} action is instead more difficult to distinguish. 
Table~\ref{tab:gmailreport} reports precision, recall and F-measure for 
different configurations of weights and packets intervals 
constraints. We can observe that the action \emph{open chats} 
(that allows to read past chats) achieves a 
low precision but a high recall. Analyzing the confusion matrix depicted in 
Figure ~\ref{fig:gmailconfusion} it is possible to notice that 
$16\%$ of \emph{other} actions wrongly classified as \emph{open 
chats}. This is the reason of such a low precision.

\begin{table}[t]
\begin{center} {%\footnotesize
\begin{tabular}{|l|r|r|r|}
    \hline
%  \multicolumn{5}{c}{Gmail with short 0 (Conf. 1)weight set and cut 0.6}  \\
    \textbf{Actions} & \textbf{Precision}& \textbf{Recall} & 
\textbf{F-measure}\\
    \hline
 \emph{send mail} &      1.00   &   1.00 &     1.00 \\
\emph{reply button}    &   0.85    &  1.00   &   0.92 \\
  \emph{open chats}    &   0.36    &  0.94  &    0.52 \\
\emph{send reply} &      0.98   &   1.00   &   0.99 \\
\emph{other}   &    0.99  &    0.82   &   0.90  \\
\hline
Average    &   0.83   &   0.85  &    0.86  \\ %arithmetic mean
    \hline
\end{tabular} }
\end{center}
\caption{Classification results of Gmail actions reached by using {Configuration~1}.}
\label{tab:gmailreport}
\end{table}

\subsubsection{Twitter}
\label{ClassPerf:Twitter}
During the analysis we noticed that Twitter actions may be more 
difficult to classify than Gmail and Facebook actions. Indeed, different 
Twitter actions generate similar time series that have in common a large 
portion. Only the last three or four packets of each time series show some 
difference. Nevertheless, we have been able to reach outstanding results also 
for this app. In particular, we focus on six specific user actions: \emph{refresh 
home}, \emph{open contacts}, \emph{tweet/message},  \emph{open messages}, 
\emph{open twitter}, \emph{open tweets}. 

On average, the F-measure is equal 
to 97\%, with a precision and a recall of 98\% and 97\% respectively (see 
Table~\ref{tab:twitterreport}). 
Performance reached are reported in Figure~\ref{fig:twitterperf}. 
For each action at least one of the configurations exceeds 96\% of 
accuracy, %TODO check!!!
while the worst configuration has an accuracy in any case higher than 91\%.
The action \emph{open twitter} has accuracy and recall equal to  
100\%, independently of the Configuration set used for the clustering phase. As a 
consequence, none of examples of the test set have been wrongly classified. 
Figure~\ref{fig:twitterconfusion} reports the confusion matrix obtained by 
considering the Twitter actions.
Three of the six analyzed actions are correctly classified in more than the 99\% of the cases. 
However, the other three actions, \emph{open contacts}, \emph{open messages} and \emph{open tweets} are correctly classified in more than 95\% of the cases.
%Unfortunately, this action is classified as \emph{other} in 17\% of the cases. 
 
\begin{table}[t]
\begin{center} {%\footnotesize
\begin{tabular}{|l|r|r|r|}
  \hline
%  \multicolumn{5}{c}{Twitter with conf. 1 weight set and cut at 1.1}  \\
 \textbf{Actions} & \textbf{Precision}& \textbf{Recall} & \textbf{F-measure}\\
  \hline
  %twitter TWA6_longr1_tw-4_s7 1.1 RandomForestClassifier_60 
\emph{refresh home} & 0.94 & 0.99 & 0.96 \\ 
\emph{open contacts} & 0.97 & 0.96 & 0.97 \\ 
\emph{tweet/message} & 0.97 & 1.00 & 0.98 \\ 
\emph{open messages} & 1.00 & 0.95 & 0.97 \\ 
\emph{open twitter} & 1.00 & 1.00 & 1.00 \\ 
\emph{open tweets} & 1.00 & 0.95 & 0.97 \\
\emph{other} & 0.96 & 0.96 & 0.96 \\ 
\hline
Average & 0.98 & 0.97 & 0.97 \\  
  \hline
\end{tabular} }
\end{center}
\caption{
Classification results of Twitter actions reached 
by using the {Configuration~1}.}
\label{tab:twitterreport}
\end{table}

%%%%%%%%%%%%%%%%%%%%%%%%%%%%%%%%%%%%%

% \begin{figure}
%         \centering
%         \begin{subfigure}[t]{0.45\textwidth}
%               \centering
% \epsfig{
% file=facebook_RandomForestClassifier_60_TWA6_longr0_tw-3_s7_cut_07.png,
% width=8cm}
%   \caption{Facebook actions confusion matrix}
%   \label{fig:facebookconfusion}  
%         \end{subfigure}%
%         \\
%          %add desired spacing between images, e. g. ~, \quad, \qquad etc.
%           %(or a blank line to force the subfigure onto a new line)
%         \begin{subfigure}[t]{0.45\textwidth}
%                 \centering
% \epsfig{
% file=gmail_RandomForestClassifier_60_TWA6_short0_tw-3_s6_cut_06.png,
% width=8cm}
%     \caption{Gmail actions confusion matrix}
%     \label{fig:gmailconfusion}    
%         \end{subfigure}
%         \\
%          %add desired spacing between images, e. g. ~, \quad, \qquad etc.
%           %(or a blank line to force the subfigure onto a new line)
%         \begin{subfigure}[t]{0.45\textwidth}
%       \centering
% \epsfig{
% file=twitter_RandomForestClassifier_60_TWA6_longr1_tw-4_s7_cut_04.png,
% width=8cm}
%   \caption{Twitter actions confusion matrix}
%   \label{fig:twitterconfusion}           
%         \end{subfigure}
%          %add desired spacing between images, e. g. ~, \quad, \qquad etc.
%           %(or a blank line to force the subfigure onto a new line)
%         \caption{Confusion Matrices of Facebook, Gmail and Twitter actions.
% }\label{fig:confusionMatrices}
% \end{figure}

%%%%%%%%%%%%%%%%%%%%%%%%%%%%%%%%%%%%%  

%\newpage
\section{Possible Countermeasures}
\label{Countermeasures}

Users and service providers might believe that their two parties communications 
are secure if they use the right encryption and authentication mechanisms. 
Unfortunately, current secure communication mechanisms limit their 
traffic encryption actions to the syntax of the transmitted data. The semantic 
of the communication is not protected in any 
way~\cite{Krishnamurthy:2013:POS:2498345.2498600}. For this reason, it has been 
possible for example to develop classifiers for TLS/SSL encrypted traffic 
that are able to discriminate between applications.

The contribution of this paper was to investigate to which extent it is feasible to identify the specific
actions that a user is doing on his mobile device, by simply eavesdropping the device's network traffic.
While it is out of the scope of the paper to investigate possible countermeasure to the proposed attack, we discuss in the following some related issues.

One common belief is that simple padding techniques may be effective against traffic analysis approaches. 
However, it has to be considered that padding countermeasures are already standardized in TLS, 
explicitly to ``frustrate attacks on a protocol that are based on
analysis of the lengths of exchanged messages''~\cite{rfc5246}. 
Nevertheless, our attack worked against TLS encrypted traffic.
More advanced techniques have been proposed in the literature, such as traffic morphing and direct target sampling \cite{conf/ndss/WrightCM09,Wright:2008:SMY:1397759.1398055}.
However, a recent result showed that none of the existing countermeasures are effective~\cite{Dyer:2012:PIS:2310656.2310689}.
The intuition is that coarse information is unlikely to be hidden efficiently, and the analysis of 
these features may still allow an accurate analysis. 
On the light of these results, we believe it is not trivial to propose effective countermeasures to the attack we shown in this paper.
%it is out of the scope of this paper to 
%find and evaluate an effective countermeasure for the proposed attack, 
%(or for the more general problem of the traffic analysis.
Indeed, it is intention of the authors to highlight a problem that is becoming even more
alarming after the revelation about the mass surveillance programs that are nowadays
adopted by governments and nation states. 

%\newpage
%%%%%%%%%%%%%%%%%%%%%%%
\section{Conclusions}
\label{Conclusions}

We proposed a framework to analyze encrypted network traffic and to infer which particular actions the 
user executed on some apps installed on his mobile-phone. 
\hilight{We demonstrated that despite the use of SSL/TLS, our traffic analysis 
approach is an effective tool that an eavesdropper 
can leverage to undermine the privacy of mobile users. 
With this tool an adversary may easily learn habits of the target users. The adversary may aggregate
data of thousand users in order to gain some commercial or intelligence advantage against some competitor. In addition, a powerful attacker such as a Government, could use these insights in order to de-anonimize user actions that may be of particular interest.}
We hope that this work will \hilight{shed light on} the possible attacks that may undermine the 
user privacy, and that it will stimulate researchers to work on efficient countermeasures that can be
adopted also on mobile devices. These countermeasures may require a kind of trade-off between 
power efficiency and the required privacy level.

% \subsection{Future work}
% \label{FutureWork}
% Classification could also evaluate which clusters are relevant to classify an user action.
% Starting from those clusters, so flows inside them, it'll be possible to do other analysis to understand the meaning for every packet inside a flow.
% This kind of analysis could, for example, evaluate the exact dimension of content for post, messages or mails.
% We intent to procede our work trying to classify network traffic generated by user activity on other Android Application.
% In our work we consider only traffic directly related with user interaction, a possible future work
% could be considering the ``\emph{other side of the coin}''. 
% This mean classify traffic related to receiving mail,messages or notification from actions made by other users.
% 
% Another possible one, it could be replicate our framework on network traffic generates by the same apps on iOS devices.

% %%%%%%%%%%%%%
  \section{Acknowledgments}
  \label{Acknowledgement}
  %Mauro Conti is supported by a EU Marie Curie Fellowship for the project PRISM-CODE (grant n. PCIG11-GA-2012-321980). This work has been partially supported by the TENACE PRIN Project (grant n. 20103P34XC) funded by the Italian MIUR. 
 \hilight{Mauro Conti is supported by a Marie Curie Fellowship funded by the
European Commission under the agreement n. PCIG11-GA-2012-321980.
This work has been partially supported by the TENACE PRIN Project
20103P34XC funded by the Italian MIUR, and by the Project ``Tackling
Mobile Malware with Innovative Machine Learning Techniques'' funded by
the University of Padua.}
  We would like to thank Fabio Aiolli, Michele Donini, and Mauro Scanagatta for their insightful comments.

%%%%%%%%%%%%%
\bibliographystyle{abbrv}
\balance
\bibliography{bibliography_small}

\begin{thebibliography}{10}

\bibitem{androidRank}
Androidrank.
\newblock \url{http://www.androidrank.org/}.

\bibitem{scikitlearn}
Scikit-learn: Machine learning in {P}ython.
\newblock \url{http://scikit-learn.org/stable/}.

\bibitem{ebizmbaRank}
Top 15 most popular social networking sites, {M}ay 2014.
\newblock \url{http://www.ebizmba.com/articles/social-networking-websites}, May
  2014.

\bibitem{iphonetrackposition:14}
R.~Abir.
\newblock iphone 5s can track user’s every move even after the battery dies.
\newblock
  \texttt{http://guardianlv.com/2014/03/iphone-5s-can-track-users-every-move-even-after-the-battery-dies/},
  Mar. 2014.

\bibitem{LeoneConti.6732964}
C.~A. Ardagna, M.~Conti, M.~Leone, and J.~Stefa.
\newblock An anonymous end-to-end communication protocol for mobile cloud
  environments.
\newblock {\em Services Computing, IEEE Transactions on}, 2014.

\bibitem{Atterer:2006:KUM:1135777.1135811}
R.~Atterer, M.~Wnuk, and A.~Schmidt.
\newblock Knowing the user's every move: User activity tracking for website
  usability evaluation and implicit interaction.
\newblock In {\em Proceedings of WWW}, 2006.

\bibitem{Benevenuto:2012:CUN:2169463.2169583}
F.~Benevenuto, T.~Rodrigues, M.~Cha, and V.~Almeida.
\newblock Characterizing user navigation and interactions in online social
  networks.
\newblock {\em Inf. Sci.}, July 2012.

\bibitem{Berthold:2000:PLU:332186.332211}
O.~Berthold, H.~Federrath, and M.~K\"{o}hntopp.
\newblock Project anonymity and unobservability in the internet.
\newblock In {\em Proceedings of CFP}, 2000.

\bibitem{Breiman:2001:RF:570181.570182}
L.~Breiman.
\newblock Random forests.
\newblock {\em Machine Learning}, 45, 2001.

\bibitem{cai2012touching}
X.~Cai, X.~C. Zhang, B.~Joshi, and R.~Johnson.
\newblock Touching from a distance: Website fingerprinting attacks and
  defenses.
\newblock In {\em Proceedings of CCS}, 2012.

\bibitem{shuosp2010}
S.~Chen, R.~Wang, X.~Wang, and K.~Zhang.
\newblock Side-channel leaks in web applications: A reality today, a challenge
  tomorrow.
\newblock In {\em Proceedings of S\&P}, 2010.

\bibitem{DBLP:ContiDG13mithys}
M.~Conti, N.~Dragoni, and S.~Gottardo.
\newblock Mithys: Mind the hand you shake-protecting mobile devices from ssl
  usage vulnerabilities.
\newblock In {\em Security and Trust Management}. 2013.

\bibitem{DaiTWNS:2013}
S.~Dai, A.~Tongaonkar, X.~Wang, A.~Nucci, and D.~Song.
\newblock Networkprofiler: Towards automatic fingerprinting of android apps.
\newblock In {\em Proceedings of INFOCOM}, 2013.

\bibitem{rfc5246}
T.~Dierks and E.~Rescorla.
\newblock {The Transport Layer Security (TLS) Protocol Version 1.2}.
\newblock RFC 5246 (Proposed Standard), August 2008.

\bibitem{Dingledine:2004:TSO:1251375}
R.~Dingledine, N.~Mathewson, and P.~Syverson.
\newblock Tor: The second-generation onion router.
\newblock In {\em Proceedings of SSYM}, 2004.

\bibitem{Dyer:2012:PIS:2310656.2310689}
K.~P. Dyer, S.~E. Coull, T.~Ristenpart, and T.~Shrimpton.
\newblock Peek-a-boo, {I} still see you: Why efficient traffic analysis
  countermeasures fail.
\newblock In {\em Proceedings of S\&P}, 2012.

\bibitem{Enck:2010:TIT:1924943.1924971}
W.~Enck, P.~Gilbert, B.-G. Chun, L.~P. Cox, J.~Jung, P.~McDaniel, and A.~N.
  Sheth.
\newblock Taintdroid: An information-flow tracking system for realtime privacy
  monitoring on smartphones.
\newblock In {\em Proceedings of OSDI}, 2010.

\bibitem{Fahl:2012:WEM:2382196.2382205}
S.~Fahl, M.~Harbach, T.~Muders, L.~Baumg\"{a}rtner, B.~Freisleben, and
  M.~Smith.
\newblock Why eve and mallory love android: An analysis of android ssl
  (in)security.
\newblock In {\em Proceedings of CCS}, 2012.

\bibitem{Falaki:2010:FLT:1879141.1879176}
H.~Falaki, D.~Lymberopoulos, R.~Mahajan, S.~Kandula, and D.~Estrin.
\newblock A first look at traffic on smartphones.
\newblock In {\em Proceedings of IMC}, 2010.

\bibitem{Georgiev:2012:MDC:2382196.2382204}
M.~Georgiev, S.~Iyengar, S.~Jana, R.~Anubhai, D.~Boneh, and V.~Shmatikov.
\newblock The most dangerous code in the world: Validating ssl certificates in
  non-browser software.
\newblock In {\em Proceedings of CCS}, pages 38--49, 2012.

\bibitem{Go:2013:TAA:2444776.2444779}
Y.~Go, D.~F. Kune, S.~Woo, K.~Park, and Y.~Kim.
\newblock Towards accurate accounting of cellular data for tcp retransmission.
\newblock In {\em Proceedings of HotMobile}, 2013.

\bibitem{hastie09statisticallearning}
T.~Hastie, R.~Tibshirani, and J.~Friedman.
\newblock {\em The Elements of Statistical Learning (2nd ed.)}.
\newblock Springer New York Inc., 2009.

\bibitem{Herrmann:2009:WFA:1655008.1655013}
D.~Herrmann, R.~Wendolsky, and H.~Federrath.
\newblock Website fingerprinting: Attacking popular privacy enhancing
  technologies with the multinomial naive-bayes classifier.
\newblock In {\em Proceedings of CCSW}, 2009.

\bibitem{fbtwturkey2}
S.~Hutchinson.
\newblock Social media plays major role in turkey protests.
\newblock \url{http://www.bbc.com/news/world-europe-22772352}, June 2013.

\bibitem{Krishnamurthy:2013:POS:2498345.2498600}
B.~Krishnamurthy.
\newblock Privacy and online social networks: Can colorless green ideas sleep
  furiously?
\newblock {\em IEEE Security and Privacy}, 2013.

\bibitem{Liberatore:2006:ISE:1180405.1180437}
M.~Liberatore and B.~N. Levine.
\newblock Inferring the source of encrypted http connections.
\newblock In {\em Proceedings of CCS}, 2006.

\bibitem{gmailbeatshotmail}
S.~Ludwig.
\newblock Gmail finally blows past hotmail to become the world’s largest
  email service.
\newblock
  \url{http://venturebeat.com/2012/06/28/gmail-hotmail-yahoo-email-users/},
  June 2012.

\bibitem{Luo11httpos:sealing}
X.~Luo, P.~Zhou, E.~W.~W. Chan, W.~Lee, R.~K.~C. Chang, and R.~Perdisci.
\newblock Httpos: Sealing information leaks with browser-side obfuscation of
  encrypted flows.
\newblock In {\em Proceedings of NDSS}, 2011.

\bibitem{mitchell1997machine}
T.~M. Mitchell.
\newblock {\em Machine learning.}
\newblock 1997.

\bibitem{Muller:2007:IRM:1324818}
M.~M\"{u}ller.
\newblock {\em Information Retrieval for Music and Motion}.
\newblock Springer-Verlag New York, Inc., 2007.

\bibitem{Panchenko:2011:WFO:2046556.2046570}
A.~Panchenko, L.~Niessen, A.~Zinnen, and T.~Engel.
\newblock Website fingerprinting in onion routing based anonymization networks.
\newblock In {\em Proceedings of WPES}, 2011.

\bibitem{tor:websites:fingerprint}
M.~Perry.
\newblock Experimental defense for website traffic fingerprinting.
\newblock
  \texttt{https://blog.torproject.org/blog/experimental-defense-website-traffic-fingerprinting},
  Sept. 2011.

\bibitem{Raymond:2001:TAP:371931.371972}
J.-F. Raymond.
\newblock Traffic analysis: Protocols, attacks, design issues, and open
  problems.
\newblock In {\em Designing Privacy Enhancing Technologies}, 2001.

\bibitem{InfFlow-TIFS}
B.~P. Rocha, M.~Conti, S.~Etalle, and B.~Crispo.
\newblock Hybrid static-runtime information flow and declassification
  enforcement.
\newblock {\em Information Forensics and Security, IEEE Transactions on}, 2013.

\bibitem{schlegel2011soundcomber}
R.~Schlegel, K.~Zhang, X.-y. Zhou, M.~Intwala, A.~Kapadia, and X.~Wang.
\newblock Soundcomber: A stealthy and context-aware sound trojan for
  smartphones.
\newblock In {\em Proceedings of NDSS}, 2011.

\bibitem{Schneider:2009:UOS:1644893.1644899}
F.~Schneider, A.~Feldmann, B.~Krishnamurthy, and W.~Willinger.
\newblock Understanding online social network usage from a network perspective.
\newblock In {\em Proceedings of IMC}, 2009.

\bibitem{Song:2001:TAK:1251327.1251352}
D.~X. Song, D.~Wagner, and X.~Tian.
\newblock Timing analysis of keystrokes and timing attacks on ssh.
\newblock In {\em Proceedings of SSYM}, pages 25--25, 2001.

\bibitem{angelaTrack:14}
C.~Staff.
\newblock Germany: U.s. might have monitored merkel's phone.
\newblock
  \url{http://edition.cnn.com/2013/10/23/world/europe/germany-us-merkel-phone-monitoring/},
  Oct. 2014.

\bibitem{Stober:2013:YSY:2462096.2462099}
T.~St\"{o}ber, M.~Frank, J.~Schmitt, and I.~Martinovic.
\newblock Who do you sync you are?: Smartphone fingerprinting via application
  behaviour.
\newblock In {\em Proceedings of WiSec}, 2013.

\bibitem{survey13suarez}
G.~Suarez-Tangil, J.~E. Tapiador, P.~Peris, and A.~Ribagorda.
\newblock Evolution, detection and analysis of malware for smart devices.
\newblock {\em IEEE Communications Surveys \& Tutorials}, 2013.

\bibitem{NinoVerdeNATleftBehind}
N.~V. Verde, G.~Ateniese, E.~Gabrielli, L.~V. Mancini, and A.~Spognardi.
\newblock No nat'd user left behind: Fingerprinting users behind nat from
  netflow records alone.
\newblock In {\em Proceedings of ICDCS}, 2014.

\bibitem{Wei:2012:PMP:2348543.2348563}
X.~Wei, L.~Gomez, I.~Neamtiu, and M.~Faloutsos.
\newblock Profiledroid: Multi-layer profiling of android applications.
\newblock In {\em Proceedings of Mobicom}, 2012.

\bibitem{fbtwarabstring}
D.~Wolman.
\newblock Facebook, twitter help the arab spring blossom.
\newblock \url{http://www.wired.com/2013/04/arabspring/}, Apr. 2013.

\bibitem{Wright:2008:SMY:1397759.1398055}
C.~V. Wright, L.~Ballard, S.~E. Coull, F.~Monrose, and G.~M. Masson.
\newblock Spot me if you can: Uncovering spoken phrases in encrypted voip
  conversations.
\newblock In {\em Proceedings of S\&P}, 2008.

\bibitem{conf/ndss/WrightCM09}
C.~V. Wright, S.~E. Coull, and F.~Monrose.
\newblock Traffic morphing: An efficient defense against statistical traffic
  analysis.
\newblock In {\em Proceedings of NDSS}, 2009.

\bibitem{MOSES-TPDS}
Y.~Zhauniarovich, G.~Russello, M.~Conti, B.~Crispo, and E.~Fernandes.
\newblock Moses: Supporting and enforcing security profiles on smartphones.
\newblock {\em Dependable and Secure Computing, IEEE Transactions on}, 2014.

\bibitem{zhouccs2013}
X.~Zhou, S.~Demetriou, D.~He, M.~Naveed, X.~Pan, X.~Wang, C.~A. Gunter, and
  K.~Nahrstedt.
\newblock Identity, location, disease and more: Inferring your secrets from
  android public resources.
\newblock In {\em Proceedings of CCS}, 2013.

\end{thebibliography}
\end{document}